\documentclass[]{aastex701}

\usepackage{multirow}

\newcommand{\new}[1]{\noindent\color{black}{#1}}

\shorttitle{Gravitational Redshift}
\shortauthors{Jakab \& Morsink}

\begin{document}

\title{Gravitational Redshift for Rapidly Rotating Neutron Stars}


\author{Zsombor Jakab}
\email{zjakab@ualberta.ca}
\affiliation{Department of Physics, University of Alberta, Edmonton, AB, T6G 2E1, Canada}
\author{Sharon M. Morsink}
\email{morsink@ualberta.ca}
\affiliation{Department of Physics, University of Alberta, Edmonton, AB, T6G 2E1, Canada}

\begin{abstract}

The Oblate Schwarzschild (OS) approximation is a method often used to compute the flux of X-rays emitted from a rapidly rotating neutron star. In this approximation, the oblate shape of the rotating star is embedded in the Schwarzschild metric, which is used to compute the redshift of photon energies as they propagate from the star to the telescope. In this paper, we demonstrate that there are small errors introduced by the standard treatment of photon redshift in the OS approximation and provide a simple method to correct these errors. These errors are constant in phase, so this results in a constant absolute reduction in the flux.  For PSR J0740+6620, the most rapidly spinning of the pulsars observed by NICER, we estimate the flux errors are less than 1\%, which is an order of magnitude smaller than the uncertainty in the distance, so this does not affect the mass and radius constraints found for this pulsar. The errors for the other pulsars observed by NICER are even smaller. However, this correction should be included when analyzing data for more rapidly rotating X-ray pulsars with spin frequencies near 600 Hz.

\end{abstract}


\keywords{Neutron stars (1108), Millisecond pulsars (1062), Rotation powered pulsars (1408), Stellar rotation (1629), Relativistic stars (1392)}


\section{Introduction} \label{sec:intro}

The equation of state (EOS) of the cold, supranuclear density 
matter in the cores of neutron stars is still unknown. Since each proposed EOS can be mapped to a unique mass-radius curve, with enough measurements of the masses and radii of neutron stars, the correct EOS could, in principle, be determined \citep{1992Lindblom}. While the masses of some neutron stars can be measured if they are in a binary system, determinations of the radius require model-dependent methods.

One method for estimating the radius is through the observation of X-rays emitted from the neutron star's surface. Photons travel to the observer on curved paths that sample the neutron star's gravitational field. Additionally, the photon's energy changes due to the change in the gravitational potential and the relative motion between the emitting region and the observer. These relativistic effects leave signatures in the observed X-rays that can allow estimations of the neutron star's mass and radius. This type of methodology has been used to constrain the masses and radii through spectral timing observations of pulsed emission from spots by the NICER X-ray telescope \citep{2019bBogdanov}. Observations of the X-ray spectrum of unpulsed emission by the Chandra and XMM-Newton telescopes have also been used to constrain the masses and radii of some neutron stars in quiescent low-mass X-ray binaries (qLMXBs) \citep{2018Steiner} and type I X-ray burst sources \citep{2017A&A...608A..31Nattila}.

Given a set of neutron star properties, such as spin frequency, mass, radius, atmosphere model, observer inclination angle, and a set of spot sizes, shapes, locations, and temperatures, it is possible to solve the relativistic stellar structure equations and geodesic equations to predict the X-ray flux that would be measured by an observer \citep{2007ApJ...654..458Cadeau}. However, the first principle construction of the X-ray flux as done by \cite{2007ApJ...654..458Cadeau} is computationally expensive, so approximation schemes are introduced to make the inverse problem (the determination of the star's parameters from the observed flux) tractable. The simplest approximation scheme is the Schwarzschild plus Doppler (SD) approximation which includes the Doppler shifts caused by the star's rotation \citep{1998ApJ...499L..37Miller,2003Poutanen}. In the SD approximation, the gravitational light-bending and redshift are included using the Schwarzschild metric as though the star were not rotating. Meanwhile, the SD approximation includes the most important rotational effects: the Doppler shift and beaming. 


\citet{2007ApJ...654..458Cadeau} showed that the SD approximation is a good approximation for very slow rotation, but that the deformation of the surface into an oblate spheroid for moderate rotation rates affects the observed flux. This motivated the development of the Oblate Schwarzschild (OS) approximation \citep{2007Morsink} which takes into account the changes in the allowed initial photon directions caused by the non-spherical shape by embedding the oblate shape in the Schwarzschild metric, leading to an easily implemented computationally-cheap approximation that is similar to the SD approximation with the same photon deflections, and gravitational and Doppler shifts. Further analysis \citep{2019bBogdanov} has shown that the correction terms due to the OS are large enough that they should be included in the analysis of the rotation-powered pulsars (that spin with rates $\sim 200 - 350 $Hz) observed by NICER.  

The OS approximation has three main known sources of errors. The first is the approximation used to define the oblate shape (for example \citet{2021PhRvD.103f3038Silva,2025PhRvD.111h3056Papigkiotis}). The second is the change in photon trajectories caused by the angular momentum of the spacetime. Since the deflection of light rays is dominated by the monopole term of the gravitational potential, the Schwarzschild metric continues to be a surprisingly good approximation \citep{2018A&A...615A..50Nattila,2018ApJ...863....8Pihajoki}. 
However, it has been shown \citep{2021MNRAS.505.2870Oliva} that the non-spherically symmetric contributions to the gravitational field, such as frame-dragging and the quadrupole moment, can contribute to changes in the geodesics and the predicted flux for rapid spin rates above 600 Hz. 
We intend to examine both of these sources of error in more detail in future work.

The third source of error in the OS approximation is introduced by the change in the photon's energy as it travels from the star to the observer. The ratio of the observed to emitted photon energies is  $E_{obs}/E_{em} = (1+z)^{-1}$, where $z$ is commonly known as the redshift.
The redshift is particularly important since the observed specific intensity is proportional to $(1+z)^{-3}$, so small errors in the redshift cause larger errors in the observed flux.
It is well known \citep{1971AcA....21...81Abramowicz,2013rrs..book.....Friedman} that the redshift of zero angular momentum (ZAM) photons does not vary with latitude on the surface of a uniformly rotating neutron star. 
However, the OS approximation introduces variations of the ZAM photons' redshift with latitude. 
In this paper, we construct an improved OS approximation that enforces the constancy of the redshift for the ZAM photons. We demonstrate that the changes introduced by this improved OS approximation are negligible for the slower rotation-powered pulsars observed by NICER. However, the corrections could be significant for neutron stars that rotate more rapidly. For example, there are at least seven accretion-powered pulsars and/or X-ray burst sources \citep{2012Watts} with spin frequencies greater than 550 Hz, for which these corrections will be important. In addition, we show that if the constancy of the ZAM photon's redshift is not enforced, spurious predictions of narrow atomic features could result for rapidly rotating stars. 

The structure of this paper is as follows. In Section \ref{sec:redshift} we provide the theoretical background for describing a rotating neutron star, the definition of redshift, and its relation to the oblate surface. In Section \ref{sec:approx} we define the various approximation schemes used in the past. The improved OS approximation is introduced and defined in Section \ref{sec:ZOS}. In Section \ref{sec:ray} we provide some examples of applications to raytracing.

\section{Gravitational Redshift for Rotating Neutron Stars}
\label{sec:redshift}

\subsection{Metric, Fluid Velocity, Definition of Surface}

The metric for a stationary, axisymmetric spacetime is \citep{1971ApJ...167..359Bardeen}
\begin{eqnarray}
    ds^2 &=& - e^{2\nu} dt^2 + \bar{r}^2 \sin^{2}\theta B^2 e^{-2\nu}\left( d\phi - \omega dt\right)^2 
    + e^{2\zeta -2\nu}\left( d\bar{r}^2 + \bar{r}^2 d\theta^2\right),
    \label{eq:metric}
\end{eqnarray}
where the metric functions $\nu$, $B$, $\omega$, and $\zeta$ depend only the coordinates $\bar{r}$ (the quasi-isotropic radial coordinate) and $\theta$. 
{\new{We work in geometrized units where $G=c=1$.}}
  The radial coordinate $r$, defined by $r = \bar{r}B e^{-\nu}$, has the property that circles at constant co-latitude $\theta$ have circumference {\new{$2 \pi r \sin\theta$}}. In the limit of zero rotation, $r$ is the radial coordinate used in the Schwarzschild metric. The metric potential $\nu$ reduces to the Newtonian gravitational potential in the weak field limit. The metric potential $\omega$ is the frame-dragging term and corresponds to the angular velocity of a zero angular momentum test particle at any particular location, as measured by an observer at infinity.

\subsubsection{Fluid velocity and acceleration}

The star's angular velocity, as measured by an observer at infinity, is $\Omega$. The fluid's 4-velocity vector $u^\alpha$ is
\begin{equation}
    u^\alpha = u^t (t^\alpha + \Omega \phi^\alpha),
    \label{eq:four-vel}
\end{equation}
where $t^\alpha$ and $\phi^\alpha$ are the timelike and angular Killing vectors, and the time component of the 4-velocity, $u^t$, defined by the normalization condition $u^\alpha u^\beta g_{\alpha\beta} = -1$ is
\begin{equation}
    u^t = \gamma(v_Z)\; {e^{-\nu}},
    \label{eq:ut}
\end{equation}
with the velocity as measured by a zero angular momentum observer (ZAMO), $v_Z$   defined by
\begin{equation}
    v_Z  = (\Omega - \omega) \bar{r} \sin\theta B e^{-2\nu},
    \label{eq:vz}
\end{equation}
and the Lorentz gamma factor for any velocity $v$ is given by the usual expression
\begin{equation}
    \gamma(v) = \left( 1 - v^2 \right)^{-1/2}.
\end{equation}
In equations (\ref{eq:ut}) and (\ref{eq:vz}), the coordinates and metric functions are evaluated on the surface of the neutron star.

The 4-velocity component $u^t=dt/d\tau$ defines the transformation between time intervals as measured by an inertial observer at infinity and an observer moving with the star's fluid. In the limit of zero rotation, $u^t$ reduces to the gravitational redshift factor for the Schwarzschild metric. However, when the fluid is rotating, $u^t$, also includes the transverse Doppler factor $\gamma(v_Z)$. 

The equation of hydrostatic equilibrium for a rigidly rotating fluid is \citep{1971ApJ...167..359Bardeen}
\begin{equation}
    \frac{dp}{\epsilon(p) + p} =  d \ln(u^t),
    \label{eq:hydrostatic}
\end{equation}
where $p$ and $\epsilon(p)$ are the fluid's pressure and energy density, related by a single-parameter equation of state (independent of temperature). The equation of hydrostatic equilibrium for an isentropic, rigidly-rotating fluid states that surfaces of constant density, pressure, and $u^t$ coincide. Since the star's surface is defined as a surface of constant pressure, the time component of the 4-velocity is constant on the oblate surface and is independent of latitude \citep{1971AcA....21...81Abramowicz,1971ApJ...167..359Bardeen}.

\subsection{Definition of Gravitational Redshift}

The redshift $z$ of a photon with energy, $E_{em}$, emitted by the star's fluid and measured by an inertial observer at infinity, $E_{obs}$, is defined by
$(1+z) = E_{em}/E_{obs}= - u^\alpha \ell_\alpha$, where $\ell^\alpha$ is the 4-velocity of the emitted photon. The expression for the redshift 
of the metric defined by Equation (\ref{eq:metric}) is
\citep{2007ApJ...654..458Cadeau}
\begin{equation}
    (1+z) = u^t \left( 1 - \Omega b_z \right) = \gamma(v_Z) e^{-\nu}\left( 1 - \Omega b_z \right) ,
    \label{eq:redshift}
\end{equation}
where we use $b_z$ to denote the component of the photon's specific angular momentum in the direction of the spin axis (commonly labeled as the z-axis in Cartesian coordinates) and $u^t$ is defined by Equations (\ref{eq:ut}) and (\ref{eq:vz}). The notation used here is changed slightly from \citet{2007ApJ...654..458Cadeau} where the $b_z$ component was simply denoted $b$.  The symbol $b$ is often used to denote the photon's full angular momentum in the Schwarzschild metric, so the $z$ subscript is required for clarity.

{\new{
Although the velocity appearing in the Lorentz $\gamma$ factor in Equation (\ref{eq:redshift}) is measured in the ZAMO frame, it was shown by \citet{2007ApJ...654..458Cadeau} (their Equations (22) - (24))
that the leading order contribution to $(1-\Omega b_z)$ is just $1 - v_i \cos\xi$, where $\xi$ is the angle between the fluid velocity and the emitted photon, and
$v_i$ is the spatial 3-velocity of the fluid measured by the inertial observer, not the velocity measured by the ZAMO. 
This makes intuitive sense, since from special relativity, we would expect that the two reference frames of importance are the frame co-rotating with the observer and the inertial frame. So the velocity appearing in the longitudinal Doppler shift term should be the fluid's velocity as measured by the inertial observer.
This leads to the interpretation of the quantity $(1-\Omega b_z)$ as the analogue of the longitudinal Doppler shift in special relativity.  Meanwhile, the combination of terms $(1-\Omega b_z)\gamma(v_Z)$ is analogous to the inverse of the Doppler boost factor in special relativity.
}
}

From the definition of redshift, a photon with zero angular momentum has a redshift denoted $z_0$ and defined by
\begin{equation}
    (1+z_0) = u^t = \gamma(v_Z) e^{-\nu}.
    \label{eq:zam-z}
\end{equation}
 This leads to the simple interpretation of $u^t$ as the redshift of a zero-angular-momentum (ZAM) photon. Since $u^t$ is independent of latitude on the surface of a rotating neutron star, all photons with zero angular momentum emitted from the surface of a rotating neutron star experience the same gravitational redshift. In the case of the Schwarzschild metric, the ZAM photons travel to the observer on a trajectory with constant $\phi$. Since all photons have a redshift that is equal to the ZAM photon's redshift multiplied by the longitudinal Doppler factor, it is important that the ZAM redshift be computed accurately.

The ZAM redshift $z_0$ should be contrasted with the polar, $z_p$, forwards, $z_f$, and backwards, $z_b$ redshifts defined by \cite{1986ApJ...304..115Friedman}. The polar redshift is the redshift for a photon emitted from the star's spin pole. These photons have zero angular momentum, so $z_0=z_p$. Since $z_0$ is independent of latitude, its equality with the polar redshift holds at other latitudes, although the polar redshift is only defined at the poles. This is equivalent to the statement that the polar redshift is independent of the direction that the photon is emitted into \citep{1986ApJ...304..115Friedman}. The forwards and backwards redshifts are the redshifts of photons emitted tangent to the surface at the equator into the direction of the equatorial plane, either in the same (forward) or opposite (backwards) direction of the star's angular velocity. The values of $z_f$ and $z_b$ correspond to the redshifts of the maximum and minimum angular momentum photons from the star, and are clearly not zero angular momentum photons.

A failing of earlier ray-tracing approximations, such as the Oblate Schwarzschild approximation, is that the redshift of ZAM photons varies with latitude. In the next section, we review the treatment of the redshift in earlier approximations and introduce an improvement to the OS approximation that enforces the constancy of the ZAM photon redshift. 

\subsection{ZAM Redshift for Fiducial Stars}

For comparison purposes, we have numerically computed the properties of two fiducial stars using the program \texttt{rns}\footnote{https://github.com/rns-alberta/rns} \citep{1995Stergioulas}. The properties of the fiducial stars were chosen so that the errors introduced by the OS approximation are large enough to be seen by eye. The properties of the two stars are shown in Table \ref{tab:fiducial}. The first fiducial star is constructed using EOS CMF1 \citep{CMF1}, a hadronic EOS with moderate stiffness and a spin frequency of 490 Hz. Since the accreting ms-period X-ray pulsars typically have spin frequencies ranging from 300 - 600 Hz, the first fiducial star has properties typical of one of these neutron stars.
The second fiducial star has more extreme properties designed to be similar to the fiducial star computed by \citet{2013ApJ...766...87Baubocka}, which has a mass of $1.4 M\odot$, a radius of $10$ km, and a spin frequency of 700 Hz. The second star is constructed with \citet{1977ApJS...33..415Arnett} EOS A (which is not presently considered to be a viable EOS since its maximum mass is lower than $2 M_\odot$). However, this choice of a very soft EOS does allow for parameters similar to those chosen by \citet{2013ApJ...766...87Baubocka}.

\begin{figure}
    \centering
    \includegraphics[width=0.5\linewidth]
    {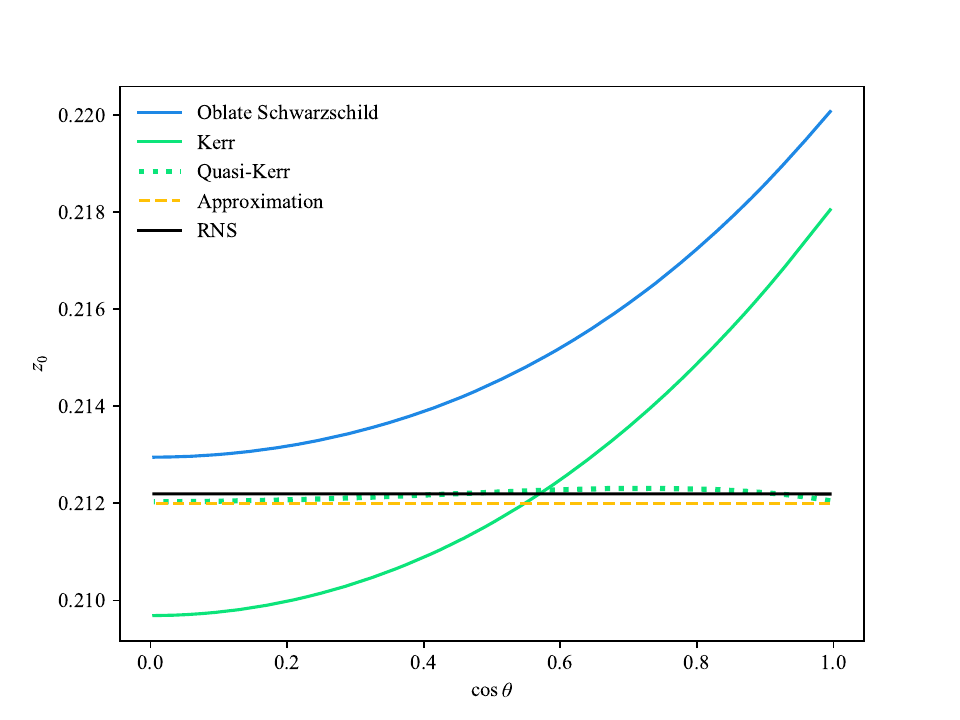}
    \includegraphics[width=0.5\linewidth]
    {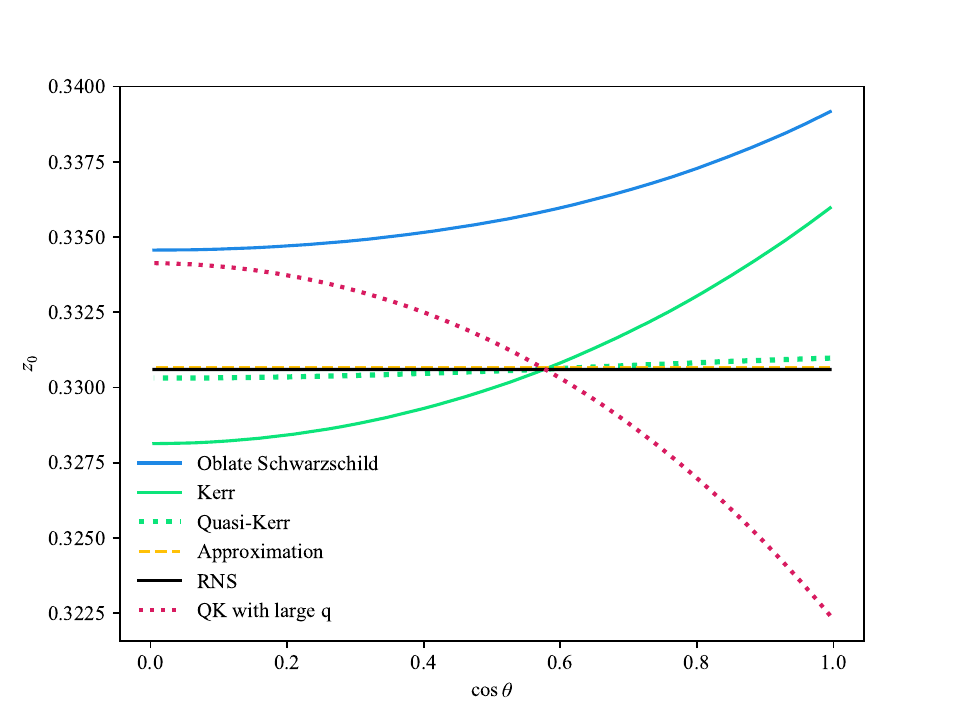}
    \caption{Redshift of ZAM photons as a function of latitude on the oblate stellar surface for the EOS DS(CMF)-1 star (upper panel) and the EOS A star (lower panel) whose properties are given in Table \ref{tab:fiducial}.
    The star's equator corresponds to 
    $\cos\theta=0$ and the spin poles are at $\cos\theta=1$.
    The solid black line is the exact value of the ZAM redshift computed using \texttt{rns} and Equation (\ref{eq:zam-z}).
    The dashed yellow line corresponds to the approximation described in Section \ref{sec:ZOS}. 
    The solid blue and green curves correspond to the Schwarzschild and Kerr metric approximations respectively. Quasi-Kerr approximations are shown with dotted curves. The green dotted curve shows the quasi-Kerr approximation when the correct values of $a$ and $q$ are used. The lower panel shows the quasi-Kerr approximation when overly large values of $a$ and $q$ are used.
    }
    \label{fig:zam}
\end{figure}

Figure \ref{fig:zam} shows plots of ZAM photon redshift as a function of latitude on the star's oblate surface for each of the two fiducial stars as a solid light black line. The ZAM photon's redshift is computed using Equation~(\ref{eq:zam-z}) at each value of latitude on the star after the star's surface has been found numerically using \texttt{rns}. Variation in the numerically computed redshift with latitude can't be seen by eye, but differs by less than ${10}^{-5}$ due to numerical error. The other curves shown on this figure represent the ZAM redshift evaluated on the same surfaces but using the approximate metrics described in the following section.

\begin{table}[]
    \centering
    \begin{tabular}{|c|c|c|}
    \hline
         & Moderate Stiffness and Speed & Soft and Fast \\
         \hline
     EOS    & DS(CMF)-1  \cite{CMF1} & A \cite{1977ApJS...33..415Arnett}\\
     $M$      & 1.44 $M_\odot$   & 1.40 $M_\odot$\\
     $R_e$    & 14.3 km         & 9.92 km\\
     $\Omega/(2\pi)$ & 490 Hz     & 700 Hz \\
     $z_0$    & 0.2122          & 0.3306 \\
     $a/M$      & 0.3454          & 0.2659 \\
     $q$      & -0.766 & -0.224\\
     $M/R_e$  & 0.1494 & 0.2087 \\
     $\overline{\Omega}$ & 0.3794 & 0.3184 \\
     \hline
    \end{tabular}
    \caption{Properties of two fiducial stars computed using \texttt{rns}. The listed properties are: equation of state (EOS), gravitational mass ($M$), equatorial radius $R_e$, spin frequency $\Omega/(2\pi)$, redshift of ZAM photon $z_0$, dimensionless angular momentum parameter $a/M = J/M^2$, dimensionless quadrupole moment $q$, compactness ratio $M/R_e$, and dimensionless angular velocity $\bar{\Omega}$.}
    \label{tab:fiducial}
\end{table}

\section{Approximation Schemes}
\label{sec:approx}

In this section we examine the treatment of the redshift in approximation schemes that make use of three different stationary, axisymmetric vacuum metrics: the Schwarzschild, Kerr, and quasi-Kerr metrics. These metrics are chosen to approximate the exact metric of a rotating neutron star computed using a program such as {\tt{rns}} given an equation of state.

In all cases, the mass parameter $M$ in the metric is chosen to be identical to the neutron star's mass. The angular momentum parameter $a$ in the Kerr and quasi-Kerr metrics is chosen to match the neutron star. In the case of the quasi-Kerr metric, the numerically computed  quadrupole moment is also used. We use the definition for the quadrupole moment introduced by \citet{1999ApJ...512..282Laarakkers} although it is not coordinate-independent \citep{2012PhRvL.108w1104Pappas} since the differences in definitions are much smaller than the effects we are examining in this paper. The oblate surface of the star $R(\theta)$, corresponding to the surface of vanishing pressure, is computed numerically from the relativistic hydrostatic equilibrium equations and embedded in each of the three metrics. The equation for the surface is usually expressed in terms of the circumferential radial coordinate $r$. It is also possible to substitute a spherical surface with a radius that does not depend on latitude, although this will not correspond to a solution of hydrostatic equilibrium. 

Since the metric (\ref{eq:metric}) is a general axisymmetric stationary metric, equation  (\ref{eq:redshift}) for the redshift holds for the different approximate metrics, as long as the correct values for $\nu$ and $v_Z$ are used.  

\subsection{Schwarzschild Approximations}

In the SD and OS approximations, the Schwarzschild metric is used to solve the geodesic equations connecting the photon emission location on the surface of the star and the observer at infinity. The photon redshift factor in an approximation that makes use of the Schwarzschild metric is 
\begin{equation}
    (1+z)_S =  \gamma(v_S) \left(1 - \frac{2M}{R(\theta)}\right)^{-1/2}  \left(1 - b_z \Omega\right)
    \label{eq:zS}
\end{equation}
where $R(\theta)$ is the function defining the star's surface as values of the radial coordinate $r$ at each latitude. The fluid velocity (as measured by an {\new{inertial}} observer located on the star's surface) is defined by
\begin{equation}
    v_S = \Omega \frac{R(\theta)}{\sqrt{1-\frac{2M}{R(\theta)}}} \sin \theta.
    \label{eq:vs}
\end{equation}
In the SD approximation, $R(\theta)$ is independent of angle. In the OS approximation, $R(\theta)$ depends on the star's mass, equatorial radius, and angular velocity in a quasi-universal manner, almost independent of the equation of state \citep{2007Morsink}.

If the star has zero angular velocity, then since the star's radius is a sphere in the SD approximation, all ZAM photons (with $b_z=0$) emitted from different latitudes will have the same redshift. However, if the star rotates, since the star's surface fluid velocity has a $\sin\theta$ dependence, the Lorentz gamma factor and the ZAM photon redshift will depend on latitude even in the SD approximation. In the case of an oblate surface, the redshift of ZAM photons given by (\ref{eq:zS}) has an even stronger dependence on $\theta$.

The special relativistic Doppler factor is defined by
\begin{equation}
    \delta^{-1} = {\gamma(v_S) (1-v_S\cos\xi)},
\end{equation}
where $\xi$ is the angle between the initial photon direction and the fluid's velocity vector.  
In the Schwarzschild metric, the motion of photons is confined to a plane, and there is a simple relationship between the angle $\xi$ and the initial co-latitude $\theta$ and longitude $\phi$ on the star where the photon is emitted, the angle $\alpha$ between the initial photon direction and the radial direction, the observer's co-latitude $\zeta$, and the bending angle $\psi$, given by
\begin{equation}
    \cos\xi = - \frac{\sin\alpha \sin \zeta \sin\phi}{\sin\psi},
\end{equation}
and
\begin{equation}
    \cos\psi = \cos\theta \cos\zeta + \sin\theta \sin\zeta \cos \phi.
\end{equation}
The definitions of these angles are shown in
Figure 1 of \cite{2019bBogdanov} and angles $\psi$ and $\alpha$ are related through the integral relation given in the same reference. 

In the Schwarzschild metric, there is a simple identity that 
\begin{equation}
    b_z \Omega = v_S \cos\xi
\end{equation}
which can be proved either through a limiting process \citep{2007ApJ...654..458Cadeau} or by rotating the plane of the photon's motion to the star's equatorial plane. As a result, it is possible to compute the redshift of a photon with any angular momentum without ray-tracing, which is why the OS approximation is computationally inexpensive.

The redshift of ZAM photons in the OS approximation for the two fiducial stars is shown in Figure \ref{fig:zam} as solid blue curves. First, the equilibrium star is computed numerically, and $M$, the mass of the star, $\Omega$, the star's angular velocity, and the star's surface $R(\theta)$ are found. (In other words, we do not make use of any of the approximate formulae for the star's surface that exist in the literature.) At each value of $\theta$, the redshift $z_S$ for $b_z=0$ is computed using equation (\ref{eq:zS}). 

The ZAM redshifts for the OS approximation shown in Figure \ref{fig:zam} are clearly not straight horizontal lines. The difference can be understood by considering the two factors entering the redshift. The factor $(1-2M/R(\theta))^{-1/2}$ is the leading order approximation to the gravitational potential $e^{-\nu}$ which is largest at the poles and smallest at the equator due to the oblate shape. However, the next largest contribution to the potential is the quadrupole moment which reduces the redshift at the poles and increases it at the equator. The lack of a quadrupole moment in the Schwarzschild metric leads to a redshift that is too large at the poles. The other factor contributing to the redshift is the Lorentz gamma factor, which is largest at the equator and reduces to 1 at the poles. The exact expression for the fluid's velocity in Equation (\ref{eq:vz}) is always smaller than the approximation arising from the Schwarzschild metric from Equation (\ref{eq:vs}), since in the correct expression, the frame-dragging angular velocity is subtracted from the star's angular velocity. The result is that the approximate Schwarzschild Lorentz factor is larger than the exact Lorentz factor. Near the equator, the increase in the Lorentz factor dominates over the decrease in the gravitational potential. The end result is that the ZAM redshift computed using the OS approximation is too large at all latitudes, however the magnitude of the error is smallest at the equator. 

In the SD approximation, the star's shape is approximated by a sphere with a radius equal to the rotating star's equatorial radius. Since the radius of the star at the equator is the same in both the SD and OS approximations the resulting approximate redshift will be the same at the equator. At higher latitudes, the contribution from $(1-2M/R_e)^{-1/2}$ will be constant, but the contribution from the Lorentz term will decrease, leading to a smaller value of redshift at the poles in the SD approximation. We have not plotted the SD approximation on Figure \ref{fig:zam} since the SD approximation is not widely used.

\subsection{Kerr Approximation}

The Kerr metric is a better approximation than the Schwarzschild metric for a rotating neutron star. In order to use the Kerr metric as an approximation, the neutron star's angular momentum, $J$ must be computed so that the correct value of the Kerr parameter, the specific angular momentum $a=J/M$ can be used.  

In order to compare the Kerr and neutron star redshifts, the Kerr metric must first be written in the form of the general axisymmetric metric given in Equation (\ref{eq:metric}). In this section, we will use the symbol $r$ to represent the Boyer-Lindquist radial coordinate, which limits to the Schwarzschild radial coordinate in the limit of zero rotation. The relation between $r$ and the coordinate $\bar{r}$ used in the general axisymmetric metic is
\begin{equation}
    r = \bar{r} \left( 1 + \frac{(M+a)}{2\bar{r}}\right)
    \left( 1 + \frac{(M-a)}{2\bar{r}}\right).
    \label{eq:rbl}
\end{equation}

The functions appearing in the metric (\ref{eq:metric}) are
\begin{eqnarray}
  \omega_K &=& \frac{2Mar}{\Upsilon}\\
    e^{2\nu_K} & =& \frac{\Sigma \Delta}{\Upsilon} \\
    B_K &=& \frac{\Delta^{1/2}}{\bar{r}} \\
    e^{2\zeta_K} &=& \frac{\Delta \Sigma^2}{\bar{r}^2 \Upsilon},
\end{eqnarray}
where the subscript $K$ denotes the Kerr values, and the
functions $\Sigma$, $\Delta$, and $\Upsilon$ are defined by
\begin{eqnarray}
    \Delta &=& r^2 - 2Mr + a^2 \\
    \Sigma &=& r^2 + a^2 \cos^2 \theta \\
    \Upsilon &=& \Sigma (r^2+a^2) + 2 Mr a^2 \sin^2 \theta.
\end{eqnarray}

The Oblate Kerr approximation to a rotating neutron star is constructed by solving for the metric (\ref{eq:metric}) of a rotating neutron star numerically, and computing the values of $M$ and $a$, as well as the equation for the star's surface $\bar{r}(\theta)$. The star's surface in Boyer-Lindquist coordinates is then found using Equation (\ref{eq:rbl}).

The ZAM photons' redshifts can be computed by substituting in the Kerr values of the metric functions into Equation~(\ref{eq:zam-z}) at different latitudes on the star's surface. The ZAM redshift as a function of latitude on the surfaces of the two fiducial stars is shown with solid green curves on Figure \ref{fig:zam}.

Two main differences from Schwarzschild should be noted. Firstly, the Kerr metric's dimensionless quadrupole parameter is $q = - (a/M)^2$, which is smaller than the neutron star's quadrupole parameter listed in Table \ref{tab:fiducial} for both fiducial stars. As a result the ZAM redshift at the poles of the stars continues to be larger than the correct value, although the values are closer to the correct value than the Schwarzschild approximation.
The equation for the fluid's speed $v_K$,
\begin{equation}
    v_K = \left(\Omega - \omega\right) \frac{\bar{r}}{r} \frac{\Upsilon \sin\theta}{\Sigma \Delta^{1/2}}
\end{equation}
is reduced by the frame-dragging term. Since the frame-dragging metric potential is approximately $\omega = 2J/r^3$ in both the Kerr and the numerical neutron star spacetimes, the values of the Lorentz gamma factors for both spacetimes are very close. Since the Lorentz factor is very close to the correct value when the Kerr metric is used, the errors in the redshift are mainly due to the Kerr quadrupole moment being too small.

\subsection{Quasi-Kerr Approximation}

The quasi-Kerr metric was introduced by \citet{2006CQGra..23.4167Glampedakis} as part of a method for distinguishing between Kerr black holes and other compact objects through the observation of gravitational radiation. The quasi-Kerr metric is the sum of two metrics, $g_{\alpha\beta} = g_{\alpha\beta}^{K} + \epsilon h_{\alpha\beta}$ where the Kerr metric $g_{\alpha\beta}^{K}$ in Boyer-Lindquist coordinates, and the term $\epsilon h_{\alpha\beta}$  allows for the matching of the mass quadrupole contribution arising from a rotating neutron star. The quantity $\epsilon$ is defined by
\begin{equation}
    \epsilon = |q| - (a/M)^2,
    \label{eq:epsilon}
\end{equation}
where $q$ is the dimensionless quadrupole moment defined by
$q = Q/M^3$. The absolute value bars are required in the definition  of $\epsilon$ since we use definitions such that $Q$ and $q$
are both negative for oblate stars. In contrast \citet{2006CQGra..23.4167Glampedakis} define $q$ as positive and $Q$ as negative. The tensor $h_{\alpha\beta}$ is defined in \citet{2006CQGra..23.4167Glampedakis} and will not be repeated here. 

The quasi-Kerr metric is a remarkably good approximation to the exterior spacetime of a rapidly rotating neutron star, once the correct values of $M$, $a$, and $q$ are chosen. This is essentially the same as the observation that the Hartle-Thorne slow rotation approximation \citep{1967Hartle,1968Hartle} is an excellent approximation for rapidly rotating neutron stars \citep{2005MNRAS.358..923Berti}. It can also be understood as the manifestation of a ``three-hair theorem" for neutron stars \citep{2015PhRvD..92b4020Majumder}. 

The quasi-Kerr metric is a perturbative expansion in powers of $\bar{\Omega}$, with $\epsilon \sim O(\bar{\Omega}^2)$. As a result, when the quasi-Kerr metric is rewritten in the form of the general axisymmetric metric (\ref{eq:metric}), the frame-dragging potential $\omega$ is given by the same expression for the Kerr metric. The fluid velocity and Lorentz factor are the same as in the Kerr metric as well.

The value of the metric potential $\nu$ required to evaluate the redshift in the quasi-Kerr approximation is
\begin{equation}
    e^{2\nu} = \frac{\Sigma \Delta}{\Upsilon} - \epsilon \left( 1 - \frac{2M}{r}\right) \left(1 - 3 \cos^2\theta \right) {\cal{F}}_1(r),
\end{equation}
where ${\cal{F}}_1(r)$, is the function
\begin{eqnarray}
    {\cal{F}}_1(r)& =& -\frac{5}{8Mr(r-2M)} (r-M)(2M^2 + 6 Mr - 3r^2) 
    - \frac{15}{16M^2} r(r-2M) \ln\left(  \frac{r}{r-2M}  \right) ,
\end{eqnarray}
derived by \cite{2006CQGra..23.4167Glampedakis}.

The ZAM redshift evaluated using the quasi-Kerr metric with values of $M$, $a$, and $q$ for the two 
fiducial stars are plotted in Figure \ref{fig:zam} using dotted lines. Each of the different dotted curves correspond to the same value of $M$, but different $a$ and $q$ values. The green dotted curves correspond to the use of the correct values of $a$ and $q$ computed with \texttt{rns}. While the quasi-Kerr redshift is not exactly constant due to the inconsistency in the use of higher multipole moments, it does provide an excellent approximation to the ZAM redshift.

 The red dotted curve shown in the lower panel of Figure \ref{fig:zam} makes use of an approximation for the quadrupole moment 
 and angular momentum 
 used by \cite{2013ApJ...766...87Baubocka} for a star with $M=1.4 M_\odot$, $R_e = 10.0$ km, and spin frequency $\nu=700$ Hz. This star has properties similar to our EOS A fiducial star. (The two stars are not identical, due to the difficulty in finding an EOS that allows for a rotating star with these specific properties.) The values that they used were $a/M = 0.357$ and $q = - 4.3 (a/M)^2 = - 0.548$ which are larger than the values computed using \texttt{rns} shown in Table \ref{tab:fiducial}. The red dotted curve shows the variation of $z_0$ with latitude resulting from the quasi-Kerr approximation with the \cite{2013ApJ...766...87Baubocka} values for the parameters, and it can be seen that there is a large variation in $z_0$, at the level of about 4\% from the pole to the equator. We will explore the effect of this variation on the profiles of spectral lines in Section \ref{sec:broadening}

Although we have not tested the quasi-Kerr metric extensively, we have found that it provides an excellent approximation to the numerically computed metric, as long as the same values of $M$, $a$, and $q$ are used, and the metric is evaluated on the correct oblate surface. 

\section{An Improved Oblate Schwarzschild Approximation}
\label{sec:ZOS}

The OS approximation introduces an incorrect dependence of the photon redshift with latitude. In the case of the fiducial star that rotates with 490 Hz, the redshift is about 4\% too large at the poles. Since the observed specific intensity is proportional to $(1+z)^{-3}$ this would lead to an error of about 2\% in the observed flux of light emitted from the poles. On the other hand, the OS approximation is a computationally inexpensive approximation, so it would be advantageous to improve the OS approximation so that it incorporates the correct redshift. 

A simple way to correct the redshift in the OS approximation is to replace the factor of $\gamma(v_S)^{-1}(1-2M/R(\theta))^{-1/2}$ appearing in equation (\ref{eq:zS}) for the Schwarzschild redshift with a better approximation for the ZAM redshift. Since the ZAM redshift is independent of latitude, we only require the value of the redshift at one latitude, which we choose to evaluate at the equator. The equatorial value is convenient, since the OS approximation requires the value of the radius evaluated at the equator, allowing for self-consistency within the OS approximation.

We should note that it is equivalent to instead evaluate the ZAM redshift at the pole, which is the value of the polar redshift $z_p$. We are aware of one formula for a universal relationship between $z_p$ and the ratio $M/R$ for stars rotating at the break-up frequency \citep{2013A&A...552A..59Bejger}. However, since we are interested in applications to neutron stars rotating at speeds much lower than breakup, we can't make use of the previous universal formula for $z_p$.

The ZAM redshift only depends on the metric, and many studies \citep{2014PhRvL.112l1101Pappas,2014PhRvD..89l4013Yagi} have shown that the metric of a rotating neutron star can be described by quasi-universal relations that only depend on the dimensionless ratios 
$M/R_e$ and $\bar{\Omega} = (R_e^3/M)^{1/2}$. It makes sense then, to seek an approximation for the ZAM redshift that depends only on these two dimensionless ratios.

\subsection{Equations of State Library}

The construction of an empirical formula for the redshift requires a choice of EOSs. We chose 30 EOS, 20 from the CompOSE\footnote{\texttt{https://compose.obspm.fr/}} Library \citep{2015PPN....46..633Typel} and 10 randomly generated piecewise polytropes \citep{Read2009}. The EOSs from the EOS database were chosen to represent a variety of particle content and calculation types. The list of CompOSE EOS used in this paper are shown in Table \ref{tab:eos}. The piecewise polytropes were generated using values consistent with nuclear physics measurements \citep{hlps-paper,2022ApJ...934..139Konstantinou}. The piecewise polytropes in our library allow for stiffer EOSs with higher maximum masses than the tabulated EOSs.

\begin{table}[]
\centering
\caption{List of equations of state from the Compose catalog \citep{2015PPN....46..633Typel} used to construct the empirical equations. EOS Name is the identifier appearing in the Compose catalog, and the Category column gives the general type of EOS. The column $M_{max}$ is the maximum mass of a non-rotating neutron star, while $R_{1.4}$ is the radius of a non-rotating neutron star with a mass of $1.4 M_\odot$.}
    \renewcommand{\arraystretch}{1.3}
    \begin{tabular}{llccl}
        \textbf{EOS Name} & \textbf{Category} & \textbf{$M_{max}$} & \textbf{$R_{1.4}$} & \textbf{Reference} \\ 
        && $M_\odot$ & km & \\ \hline \hline
        ABHT(QMC-RMF3) & Nucleonic/Relativistic density funct. & 2.15 & 12.26 & \cite{QMCRMF3}\\
        \hline
        APR(APR) & Nucleonic/Microscopic Calculations & 2.17 & 11.33 &  \cite{1998Akmal}\\ \hline
        BBB(BHF-BBB2) & Nucleonic/Microscopic Calculations & 1.92 & 11.13 & \cite{BBB}\\ \hline
        BL(chiral) with unified crust & Nucleonic/Microscopic Calculations & 2.08 & 12.27 &  \cite{BLC} \\ \hline
          CMGO (GDFM-II) & Nucleonic/Relativistic density funct. & 2.30 & 13.79 & \cite{GDFMII}\\ \hline
        DS(CMF)-1 with crust & Hyperons/Relativistic density funct. & 2.07 & 13.57 & \\ \cline{1-4}
        DS(CMF)-2 with crust & Hyperons/Relativistic density funct. & 2.13 & 13.70 & \cite{CMF1} \\ \cline{1-4} 
        DS(CMF)-1 Hybrid with crust & Quark-hadron/Relativistic density funct. & 1.96 & 13.55  \\ \hline
         GPPVA(TW) NS unified inner crust-core & Nucleonic/Relativistic density funct. & 2.07 & 12.33 & 
          \cite{1999NuPhA.656..331Typel}\\ \hline
          KBH(QHC21\_A) & Quark-hadron/Microscopic calculations & 2.19 & 12.40 &  \cite{QHC21A} \\ \hline
             PCGS(PCSB2) & Nucleonic/Relativistic density funct. & 2.02 & 13.00 &  \cite{PCSB2}\\ \hline
        RG(KDE0v) & Nucleonic/Non-relativistic density funct. & 1.97 & 11.42 & \multirow{4}{*}{\cite{KDE0v-Rs-SK255-SKa}} \\ \cline{1-4}
        RG(Rs) & Nucleonic/Non-relativistic density funct. & 2.12 & 12.93 & \\ \cline{1-4}
        RG(SK255) & Nucleonic/Non-relativistic density funct. & 2.15 & 13.15 & \\ \cline{1-4}
        RG(SKa) & Nucleonic/Non-relativistic density funct. & 2.22 & 12.92 & \\ \hline
        
        SPG(M3) unified NS EoS & Nucleonic/Relativistic density funct. & 2.69 & 12.65 & {\cite{M3-TW}} \\ \hline
      VGBCMR(D1MStar) & Nucleonic/Non-relativistic density funct. & 2.00 & 11.71 & \cite{D1MStar}\\ \hline
        XMLSLZ(DDME-X) & Nucleonic/Relativistic density funct. & 2.56 & 13.37 & \\ \cline{1-4}
        XMLSLZ(NL3) & Nucleonic/Relativistic density funct. & 2.77 & 14.59 &  \cite{DDMEX-NL3-PKDD} \\ \cline{1-4}
        XMLSLZ(PKDD) & Nucleonic/Relativistic density funct. & 2.33 & 13.63 & \\ \hline    
    \end{tabular}
    \label{tab:eos}
\end{table}

The \texttt{rns} code was used to compute the equilibrium structure and the ZAM photon redshift for about 240 stellar models for each of the 30 EOS. 
 The ranges of the central energy density $\epsilon_c$ were chosen separately for each EOS such that the masses of the zero-spin stars would be greater than or equal to $0.9 M_\odot$. The values of the polar-to-equatorial axis ratio (in the isotropic radial coordinate) ranged from 0.996 to 0.778, which allowed for stars spinning with observable spin frequencies. The data was filtered to remove neutron stars with $M>2.4 M_\odot$, $M/R < 0.05$, and spin frequencies $>$ 900 Hz. Our goal is to remove stars that would be rotating close to the breakup frequency, since these stars are highly deformed and not typical of the more slowly rotating neutron stars observed with X-ray telescopes. 
 
 Empirical fits in the following sections were done using the full set of stars from all of the 30 EOS, as well as separately on the two different families of EOS in order to test for biases introduced by the choices of EOS.

\subsection{An Approximation for the ZAM Redshift}

The redshift of a photon with zero angular momentum emitted from the equator must have a value that approaches $(1-2M/R_e)^{-1/2}$ in the limit of zero rotation. This leads to the ansatz that the ZAM redshift should have the form
\begin{equation}
    1+z_0 = \left( 1-\frac{2M}{R_e} \right) ^{-\frac{1}{2}} \left( 1 + 
    \chi(\frac{M}{R_e},\bar{\Omega}) \right),
    \label{eq:zemp}
\end{equation}
where the rotational correction factor, $\chi$, is given by
\begin{equation}
    \chi = \sum_{i=1}^{n} \sum_{j=1}^{m} \ a_{ij} \ \left( \frac{M}{R_e} \right)^i \ \overline{\Omega}^{\ 2j},
    \label{eq:fit}
\end{equation}
and the coefficients $a_{ij}$ are numbers to be determined through an empirical fit to the photon redshifts computed in the library of neutron star models.

The coefficients in the empirical formula are found by minimizing the test function 
\begin{equation}
    \sum_{i=1}^{N} \left( z_{RNS} - z_0 \right)_i ^2,
    \label{residuals}
\end{equation}
where $z_{RNS}$ is the ZAM redshift of a stellar model computed by \texttt{rns}, 
and the sum is over all models in the dataset.

We first tried a least squares regression. However, we found that the test function has a local minimum, and the least squares regression was unable to find the true minimum.
Instead, we used the method of differential evolution \citep{DiffEv} as implemented in \texttt{SciPy} \citep{2020SciPy-NMeth}. Differential evolution is a method for optimization designed to find a global minimum of a test function over a continuous space.  Fit functions, $\chi$, with increasing values of $n$ and $m$ were fit to the data until the smallest test function was found, and adding additional parameters did not improve the fit. This occurred at $n=3$, $m=2$. Test function values for all iterations can be found in Table \ref{tab:parameters}.

\begin{table}[ht!]
    \centering
    \caption{Evolution of the test function when fitting Equation \ref{eq:fit} to the dataset with increasing number of parameters}
   \begin{tabular}{C||CCC}
        & m=1      & m=2      & m=3      \\ \hline \hline
    n=2 & 5.22\times 10^{-4} & 1.01\times 10^{-4} & 1.00\times 10^{-4} \\
    n=3 & 3.13\times 10^{-4} & 8.90\times 10^{-5} & 8.92\times 10^{-5} \\
    n=4 & 3.08\times 10^{-4} & 8.91\times 10^{-5} &          \\
    \end{tabular}
    \label{tab:parameters}
\end{table}
    
\begin{figure}
    \centering
    \includegraphics[width=0.5\textwidth]{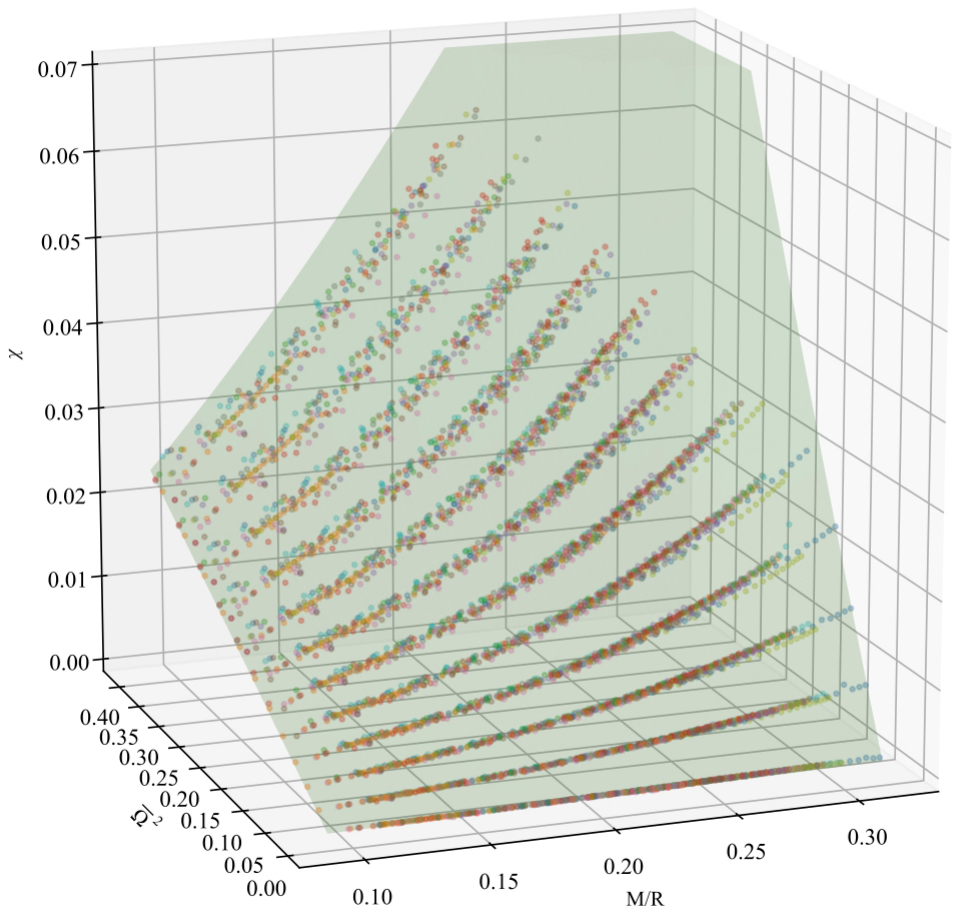}
    \includegraphics[width=0.5\textwidth]{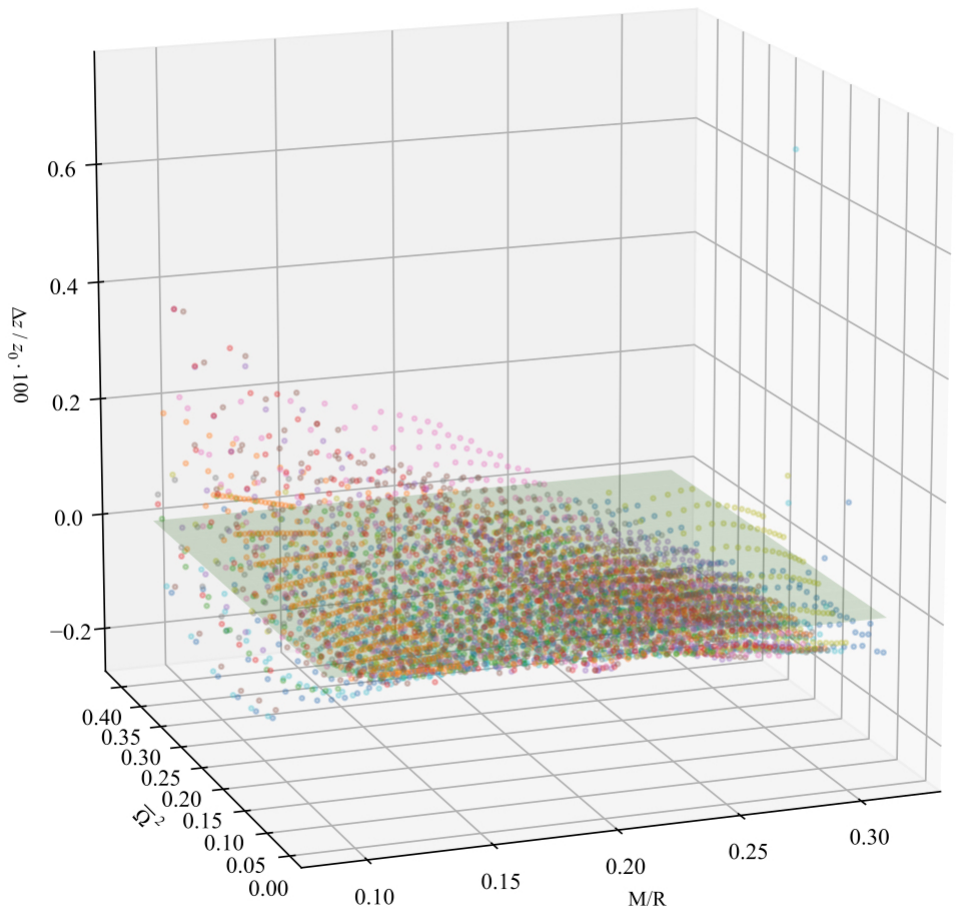} 
    \caption{ZAM redshift correction, $\chi$, (upper panel)  and the residual fractional errors (lower panel) in $z_0$ as functions of the dimensionless parameters $M/R$ and $\overline{\Omega}$. 
        The surface represents the function defined by Equation \ref{eq:fit} with $n=3$ and $m=2$. Colored points correspond to the individual calculations for each stellar model in the library.}
    \label{fig:z_surf_23}
\end{figure}

Numerical values of $a_{ij}$ for the 3x2 case are shown in Table \ref{zfit_3x2}. The correction function $\chi$ and the best fit and the corresponding percent errors for $z_0$ are shown in Figure \ref{fig:z_surf_23}. 
The largest errors are just under 0.4\%. 

\begin{table}[ht!]
\centering
    \caption{Values of $a_{ij}$ for the best redshift fit, Equation \ref{eq:fit} with $n=3$ and $m=2$.}
    \label{zfit_3x2}
    \begin{tabular}{c||ccc}
    $a_{ij}$ & i=1    & i=2 & i=3 \\ \hline \hline
    j=1 & 0.544  & 1.146                  & -1.173                 \\
    j=2 & -0.089 & 0.473                  & 1.938                  \\
    \end{tabular}
\end{table}

The best-fit coefficients shown in Table \ref{zfit_3x2} were determined using all 30 EOS. We also tried excluding the random polytrope equations of state, which led to a 0.05\% difference in the leading order parameter $a_{11}$. Since the resulting fits with and without the polytropic EOSs are almost indistinguishable, we conclude that the use of the piecewise polytropes does not bias our results. 

We also attempted fits using the dimensionless angular momentum $j$ in place of $\overline{\Omega}$ in Equation \ref{eq:fit}. This produced significantly worse fits, with residuals being about 10 times larger for the same number of parameters.

In Figure \ref{fig:zam}, the best-fit approximation to the ZAM photon redshift is represented as a dashed yellow line (labelled ``Approximation") for both fiducial stars. For both cases, the approximation given by Equation (\ref{eq:zemp}) is much closer to the actual value than the redshift approximations making use of the Schwarzschild, Kerr, or quasi-Kerr metrics. It should be noted that the 2nd fiducial star is computed using EOS A, which is an overly-soft EOS (maximum mass of $1.66 M_\odot$) that is not part of our library of EOS. This serves as a minor test that the approximate formula can be applied to neutron stars with EOS that are quite different from those in our library.

\subsection{The Improved OS Approximation}

Our proposal for an improved OS approximation is as follows. First, values of $M$, $R_e$, and $\Omega$ are chosen, and the dimensionless ratios $M/R_e$ and $\bar{\Omega}$ are computed. The surface of the star is defined by using one of the empirical shape formulae (e.g., \citet{2014AlGendy}) that accurately models the oblate shape that is the solution of the equations of hydrostatic equilibrium. The worldline of a photon connecting the emission point on the star to the observer is computed using the Schwarzschild metric for all points at and outside of the star's surface. The bending angles, for example, still depend on $M/R(\theta)$, and most other features of the OS approximation, such as the tilt of the surface normal compared to the radial direction, are unchanged from the OS approximation. 
However, the specific intensity of the emitted light is found using a new equation for the photon redshifts. In particular, the photon energy relation
\begin{equation}
    \frac{E_{obs}}{E_{em}} 
    = \frac{1}{1+z}
    =
    \frac{1}{(1+z_0)
    \left( 1 - v_S \cos\xi\right)},
    \label{eq:ZOS}
\end{equation}
should be used when evaluating the observed specific intensity, where $z_0$ is given by the approximate formula (\ref{eq:zemp}) and the fit (\ref{eq:fit}). 
In this equation, the longitudinal Doppler effect $(1-v_s \cos\xi)$ is unchanged from the Schwarzschild value since this is an exact expression for photon geodesics in Schwarzschild. We will consider the errors in the deflection angles introduced by the Schwarzschild metric elsewhere, but other computations suggest that these errors should be small \citep{2018A&A...615A..50Nattila}. 

\section{Applications}
\label{sec:ray}

The gravitational redshift plays an important role in the interpretation of observations of X-rays emitted from the surface of a neutron star. The two main applications where the photon redshift is required are the rotational broadening of spectra and pulsed emission from hot spots.

\subsection{Rotational Broadening}
\label{sec:broadening}

Stellar rotation broadens observed spectral lines due to the  Doppler effect if the light is emitted from an extended region on the star. If the photon geodesics are approximated well using the Schwarzschild metric, the observed photon energy is given by equation (\ref{eq:ZOS}). While the quantity $1+z_0$ does not vary on the surface of the star, the Schwarzschild velocity depends on $\sin\theta$ and the angle $\xi$ depends on $\theta$ and $\phi$ as well as $\zeta$, the colatitude of the observer. 

The dominant contribution to rotational broadening are the longitudinal Doppler shifts that broaden a sharp spectral line into one which covers an energy range that is roughly $\Delta E_{obs}/E_{obs} = 2 \Omega R_\infty \sin\theta \sin\zeta/c$ (where we are neglecting the $\sin\psi/\sin\alpha$ dependence caused by gravitational lensing, and $R_\infty = R/\sqrt{1-2M/R}$). As a result, large angular velocities tend to smear out spectral lines over a large range of observed energies, making them difficult to detect unless the observer's inclination is very close to the spin axis. 

Surprisingly, \citet{2013ApJ...766...87Baubocka} {\new{claimed}} that narrow spectral lines could be formed in the case of very rapid rotation and a moderate inclination angle. Our EOS A fiducial star mimics their example configuration. They attribute the narrow absorption feature to the star's quadrupole moment, which causes a claimed variation in the gravitational redshift with latitude. In 
\citet{2013ApJ...766...87Baubocka}, they plot the redshifted energy as a function of latitude on the star, for a fixed value of longitude $\phi$ in their Figure~6. In particular, the dot-dashed curve in their Figure~6 shows the redshifted energy for the quasi-Kerr approximation using their values of $a/M$ and $q$. This curve has a minimum, and as a result, a large region of the star emits photons whose detected energy is almost constant, leading to a sharp spectral line. 

The dependence of the redshifted photon energy on co-latitude $\theta$ for constant $\phi$ can be easily understood by taking the derivative of equation (\ref{eq:ZOS}) with respect to $\theta$, along a curve of constant longitude,
\begin{eqnarray}
    \frac{\partial(E_{obs}/E_{em})}{\partial \theta}
    &=& \left( 1+z \right)^{-2} \left(
    (1+z_0)\frac{\Omega R_\infty}{c} 
\frac{\partial(\sin\theta\cos\xi)}{\partial \theta}
     + \sin \theta \frac{dz_0}{d\cos\theta}
    (1 - \frac{\Omega R_\infty}{c}\sin\theta \cos\xi)
    \right).
\end{eqnarray}
The correct ZAM redshift $z_0$ is independent of latitude on the surface of a star, so the term proportional to $dz_0/d\cos\theta$ vanishes. If there is no light-bending, the partial derivative of $\cos\xi$ (for constant $\phi$ and $\zeta$) with respect to $\theta$ vanishes. 
On the blue-shifted side of the star, $\sin\phi < 0$
and $\cos\xi > 0$. As a result, the partial derivative of the observed photon energy is positive meaning that the observed photon energy increases monotonically from the pole to the equator on the blue-shifted side of the star.  In reality, gravitational lensing introduces a weak dependence of $\cos\xi$ on $\theta$, but the dependence of photon energy with $\theta$ continues to monotonically increase until the maximum energy is reached very close to the equator. If the correct ZAM redshift is evaluated on the oblate star's surface, there will be no extremum in the observed energy at high latitudes, and the spectral line will be broad.

{\new{
A rotating neutron star in hydrostatic equilibrium has a unique set of multipole moments that lead to a ZAM redshift that is constant in latitude. While the quasi-Kerr metric treats the quadrupole moment as a free parameter, choosing any arbitrary value that does not agree with hydrostatic equilibrium will lead to variations of the ZAM redshift with latitude. These incorrect variations in the redshift can lead to incorrect variations in the profiles of spectral lines.
}}
Consider the dependence of the observed energy as a function of $\theta$ if an incorrect ZAM redshift with $dz_0/d\cos\theta < 0$ as is the case for the curve labelled ``QK with large q" in Figure \ref{fig:zam}. Since this derivative is negative, if the magnitude of the derivative is large enough, it will be possible to find a value of $\theta$ at mid-latitudes where the photon energy has a minimum value, leading to a sharp peak in the photon energy. This requires that the observer's inclination angle be close to the spin axis, since $\cos\xi \sim \sin \zeta$ will suppress the longitudinal Doppler shift if $\zeta$ is small.
The non-zero derivative of $z_0$ is roughly proportional to the difference between the incorrect and correct quadrupole moments. If this difference is large enough (ie, $O( v/c \sin\zeta)$), then it will be possible to find a value of theta at mid-to-high latitudes where the observed photon has a minimum.

If the quadrupole moment is too small, as happens in the cases of the OS and Kerr approximations, the derivative $dz_0/d\cos\theta > 0$, (as can be seen in Figure \ref{fig:zam}) meaning that no minimum occurs and the energy increases monotonically from pole to equator on the blue-shifted side of the star as expected. In this case, we would expect that the resulting spectral lines will be broad. 

A few computations of rotational broadening for light emitted by neutron stars have been done in the past, none of which have seen evidence for a narrow spectral line. \citet{2003ApJ...582L..31Ozel} and \citet{2010ApJ...723.1053Lin} computed line profiles using the Schwarzschild metric and a neutron star with a spherical surface and found that spectral lines were highly broadened for rapid rotation, as would be expected. \citet{2006ApJ...636L.117Chang} computed line profiles for stars spinning up to 300 Hz, including the correct shape and quadrupole moment using the \texttt{rns} code and ray-tracing, and only found broadened spectral lines. However, that paper did not consider observer inclination angles close to the spin axis nor very rapid rotation, so they did not actually test the scenario suggested by \citet{2013ApJ...766...87Baubocka}. \citet{2006ApJ...644.1085Bhattacharyya} computed profiles using the Kerr metric and a spherical surface. They only considered emission from belts on the star's surface covering a narrow range of latitudes, and they found that narrow spectral lines could be formed if the emission is restricted to a small range of latitudes very close to the spin axis. We note that calculations restricted to a very narrow range of latitudes do not suffer from the issue of an incorrectly varying ZAM redshift since the latitude of emission does not vary. 

\citet{2018A&A...615A..50Nattila} computed line profiles using an approximate metric that self-consistently models the oblate shape, angular momentum, and quadrupole moment for rapidly rotating neutron stars spinning at 700 Hz observed at an inclination angle close to the spin axis. In particular, they found that the resulting spectral lines are broad. However, when they artificially increased the star's quadrupole moment {\new{by a factor of 4}}, they were able to reproduce the sharp spectral line found by \citet{2013ApJ...766...87Baubocka}. It is clear from the example calculations by \citet{2018A&A...615A..50Nattila} that the sharp spectral feature is due to an exaggerated quadrupole moment. {\new{Our calculations agree that a factor of 4 enhancement in the quadrupole moment will lead to a large enough variation in the ZAM redshift, providing a mechanism that explains how the incorrect sharp spectral feature is produced.}}


\subsection{Pulsed Emission}

Observations of pulsed X-ray emission from hot spots on a neutron star's surface are used to constrain the masses and radii of neutron stars using the OS approximation \citep{2019bBogdanov}. We have shown in the previous sections that the OS approximation introduces errors into the redshift and the flux calculations. In this section we examine the magnitude of these errors. 

First we consider an infinitesimal spot located at one latitude on the star. From the previous sections, we have shown that the largest error in the redshift occurs for emission close to the spin axis, 
while the smallest error occurs at the equator. The error in the redshift introduces an error in the observed specific intensity and the flux, since $I_{obs} \propto (1+z)^{-3}$. For an infinitesimal spot, the error in flux will be constant in phase, so the error introduced by the OS approximation is a constant offset in the absolute flux.

The ZAM redshift in the OS approximation can be computed using equation (\ref{eq:zS}) with $b_z=0$ and the universal formula for the 
shape of the star \citep{2014AlGendy}. Table \ref{tab:rederrors} displays the ZAM redshift evaluated at the pole and the equator in the OS approximation for a few neutron stars with representative masses, radii, and spin frequencies. The universal approximation for the ZAM redshift derived in this paper is shown in the column labelled $1+z_0$, and the percent error due to the OS approximation in $1+z$ and the specific intensity evaluated at the equator are labelled $\Delta(1+z_e)$ and $\Delta(I)_e$ respectively and similarly for the values at the pole. The values of the errors in the intensity are always negative (flux is smaller than the correct value) and the change in the error from equator to pole is monotonic.

The  rotation-powered pulsars  observed by NICER and their spin frequencies are: 174 Hz for PSR J0437-4714  \citep{2024ApJ...971L..20Choudhury}, 205 Hz for PSR J0030+0451 \citep{2019Riley,2019Miller}, 271 Hz for PSR J1231-1411 \citep{2024ApJ...976...58Salmi}, 318 Hz for PSR J0614-3329 \citep{2025arXiv250614883Mauviard}, and 347 Hz for PSR J0740+6620 
\citep{2024ApJ...974..294Salmi,2024ApJ...974..295Dittman}.
The stars in Table \ref{tab:rederrors} spinning at 200 and 300 Hz are representative of the four slower pulsars observed by NICER. For these examples, the error in the observed flux introduced by incorrect ZAM redshift used in the OS approximation is less than 0.5\% near the pole. The star spinning at 350 Hz has parameters similar to those inferred for the pulsar J0740 \citep{2024ApJ...974..295Dittman,2024ApJ...974..294Salmi}. The higher spin introduces a larger error that ranges from 0.6\% to 0.9\% from the equator to the pole.  
For all of these examples, the errors introduced by the OS approximation are tiny compared to the typical errors introduced by the other unknowns, such as the distance, surface temperature, and the counting statistics. For instance,  the distance to J0740 has an estimated uncertainty of 10\% \citep{2021ApJ...915L..12Fonseca}, which dwarfs any errors introduced by the redshift for this star.

Accreting millisecond X-ray pulsars (AMXPs), such as SAX J1808.4-3658 which spins at a frequency of 401 Hz, have much more complex properties than the rotation-powered pulsars as detailed by \citet{2025MNRAS.538.2853Dorsman}. We expect that the less than 1\% errors introduced at 400 Hz are smaller than the uncertainties introduced by the accretion physics. Similarly, the slower AMXP, XTE J1814-338 \citep{2024MNRAS.535.1507Kini}, which also has X-ray burst oscillations, rotates at 314~Hz, so we would expect that the redshift errors are swamped by other uncertainties.

A few of the AMXPs and X-ray burst oscillation sources have spin frequencies close to 600 Hz. In Table \ref{tab:rederrors}, two examples of neutron stars with spin frequencies of 600 Hz are listed, and in both cases the percent error in the flux could be as large as 3\%. At present, models of these systems, for example 4U 1636-536, which spins at 581 Hz \citep{2025MNRAS.tmp..892Kini}, are still dominated by much larger uncertainties in the accretion.
This error in redshift at 600 Hz is large enough that once systematic uncertainties in modelling accreting systems with very rapid rotation have been reduced, the improved OS approximation outlined in Section \ref{sec:ZOS} should be adopted in the codes modelling their flux.

\begin{table}[ht!]
    \centering
    \caption{Percent errors in ZAM redshift and specific intensity due to the OS approximation for a few sets of representative neutron star parameters.}
    \label{tab:rederrors}
    \begin{tabular}{lll|cc|cccc|cc|cc}
    \hline
    $M$ & $R_e$ & $\nu$ & $M/R_e$ & $\bar{\Omega}$ & $R_p$ &
    $(1+z_e)_{OS}$ & $(1+z_p)_{OS}$ & $(1+z_0)$ & 
    $\Delta(1+z_e)$ & $\Delta(I)_e$ & $\Delta(1+z_p)$ & $\Delta(I)_p$ \\
    $M_\odot$ & km & Hz &  &  & km
    & & &  & \% & \% & \% & \%
    \\ \hline
    1.4	&13.0 &	200	& 0.1590 &	0.1366 &	12.85 &	1.2136&1.2143&1.2134 & 0.01 & -0.03      &	0.07 &	-0.21 \\
    1.4	&13&	300&	0.1590&	0.2050&	12.66&	1.2169&	1.2186&	1.2166&	0.03&	-0.08&	0.17&	-0.50\\
    2.1	&13.0	&350 &	0.2385 & 0.1953 &	12.73 &	1.3950&1.3963&1.3923 & 0.19	& -0.58      &	0.29  &	-0.87 \\
    1.4 &	13&	400&	0.1590&	0.2733&	12.39&	1.2216&	1.2250&	1.2210&  0.05&	-0.15&	0.33&	-0.98\\	
1.4	&13.0&	600&	0.1590&	0.4099&	11.64&	1.2354&	1.2455&	1.2337&	0.14 &	-0.42 & 0.95&	-2.89\\
2.1&13.0&	600&	0.2385&	0.3347&	12.21&	1.4196&	1.4255&	1.4111& 0.60 &	-1.81	&1.02&	-3.09\\
\hline
    \end{tabular}
\end{table}

\begin{figure}
    \centering
    \includegraphics[width=0.5\textwidth]{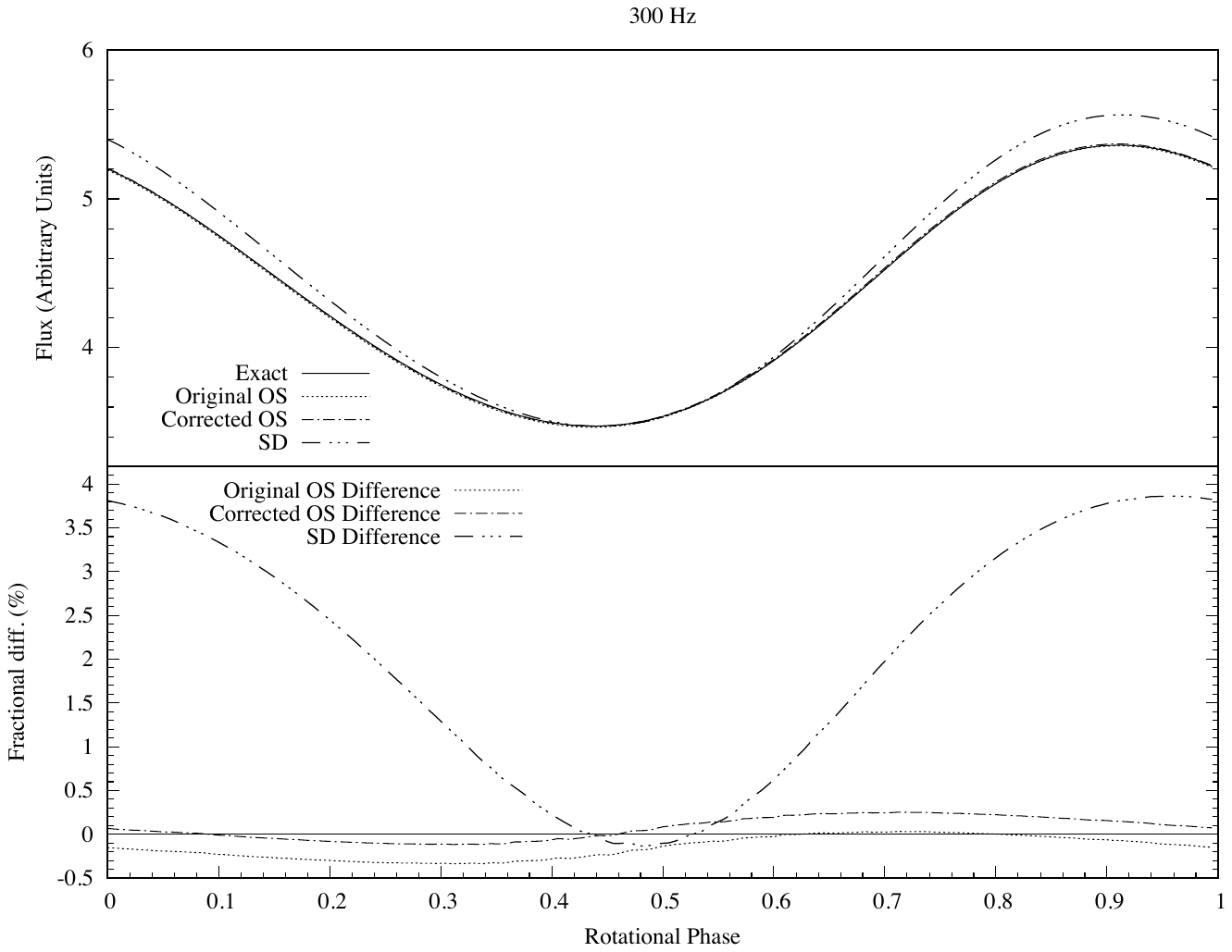}\includegraphics[width=0.5\textwidth]{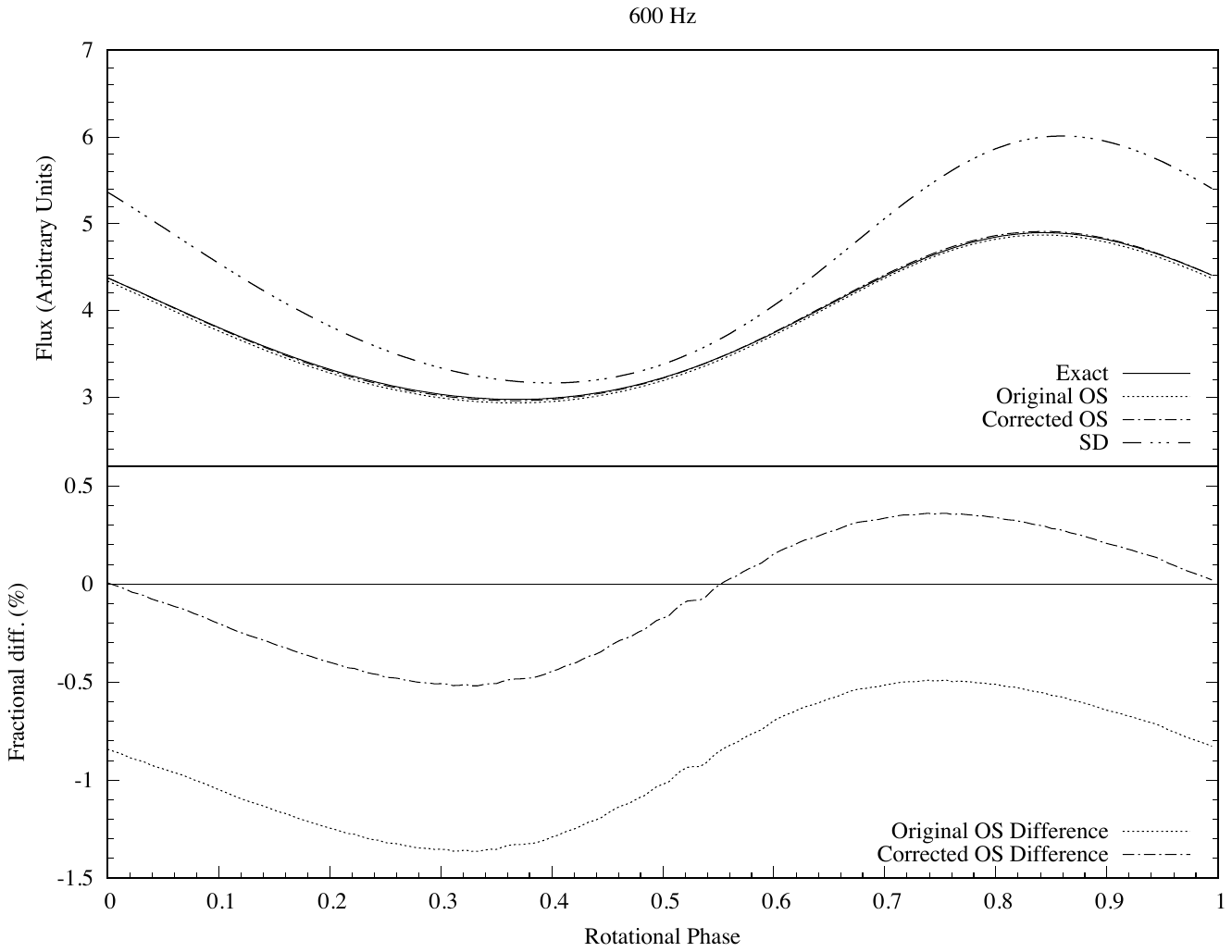}
    \caption{\new{Lightcurves (upper panels) for monochromatic light emitted from $\theta=41^\circ$ and observed at $\zeta=20^\circ$, for stars constructed with EOS L, with $M=1.4 M_\odot$ and spin frequencies of 300 Hz (left) and 600 Hz (right). The exact light curves obtained from raytracing on the correct metric \citep{2007ApJ...654..458Cadeau} are shown with solid lines. The original and corrected OS approximations are shown with dotted and dash-dotted curves. The SD approximation is shown with a dash-dotted curve with 3 dots for comparison. The percent differences between the two OS approximations with respect to the exact lightcurve is shown in the lower panel. Due to the very large errors in the SD approximation, the percent error for SD is not shown for 600 Hz. }}
    \label{fig:600Hz}
\end{figure}

{\new{As an example of the magnitude of the redshift correction, we apply the correction factor to the lightcurves constructed for the very stiff EOS~L \citep{1977ApJS...33..415Arnett}  neutron star by \citet{2007ApJ...654..458Cadeau}, with a mass of $M=1.4 M_\odot$ and frequencies of 300 and 600 Hz. The example of the 600 Hz star is illustrated in Figure 3 of \citet{2007ApJ...654..458Cadeau}. Monochromatic emission from an infinitesimal spot located at a co-latitude of 41 degrees is assumed with an observer at an inclination of 20 degrees.  We have replotted the original lightcurves (with the change that the flux is not normalized to 1) in Figure \ref{fig:600Hz}, and have added the percent difference between the exact lightcurve resulting from raytracing using the full metric \citep{2007ApJ...654..458Cadeau} and the original OS approximation, and the OS approximation with the redshift correction applied. The redshift correction provides a small upwards shift in the flux that is independent of phase. In the case of the star spinning at 300 Hz, both the original OS and corrected OS curves differ by less than 0.5\% from the exact curve. 
In the case of 600 Hz, the flux predicted by the original OS approximation was too small by as much as 1.5\%.  The redshift-corrected OS approximation is closer to the exact result, but still has a phase-dependent error of magnitude 0.5\%, which is due to the spherically-symmetric Schwarzschild metric being used to connect emitted photons to the observer, instead of the correct non-spherical axisymmetric metric.
The errors introduced when a spherical surface is used (SD) are much larger than the OS errors, and are as large as 4\% for 300 Hz, and about 20\% for 600 Hz. }}

\section{Discussion}

The equation of hydrostatic equilibrium for rotating, relativistic stars requires that the redshift of photons with zero angular momentum be independent of latitude on the star. The Oblate Schwarzschild approximation \citep{2007Morsink}, a commonly used method to model the flux of light emitted by rapidly rotating neutron stars, suffers from the issue that it introduces latitude variations in the ZAM redshift. In this paper, we examined the errors in the measured flux caused by the redshift errors in the OS approximation and introduced a simple, easily implemented correction factor that restores the constancy of the ZAM redshift.

The errors in the flux due to the redshift introduced by the OS approximation are less than 1\% for the rotation-powered pulsars observed by NICER, which is smaller than the other uncertainties in the observations. As a result, the issues discussed in this paper do not affect the mass and radius constraints found using the NICER observations. However, future observations with newer telescopes such as eXTP \citep{2025arXiv250608104Li} and NewAthena \citep{2025NatAs...9...36Cruise} will have lower statistical errors and will be able to observe higher-spin pulsars, so the modifications to the OS approximation described in this paper should be adopted in the next generation of pulse-profile models and spectral fitting programs. 

To keep the error in flux due to the redshift in context, we should consider the bigger picture of other potential errors in the OS approximation at high spin. First of all, the errors due to redshift are constant in phase and tiny compared to the phase-dependent errors that occur when the star's shape is modelled as a perfect sphere. The consideration in this paper of the errors introduced by the redshift is the first step in a larger program examining the errors introduced by the OS approximation. We plan in future work to quantify the errors in flux arising from the approximate shape functions used to model the oblate shape of the star, and the use of Schwarzschild geodesics to model the deflection of light rays for more rapidly rotating neutron stars. 

\begin{acknowledgments}
We thank Cole Miller, Shafayat Shawqi, Anna Watts, and the anonymous referee for useful comments on this work. This research was supported by NSERC
Discovery Grant RGPIN-2019-06077.
\end{acknowledgments}

\vspace{5mm}







\bibliography{biblio}{}

\begin{thebibliography}{}
\expandafter\ifx\csname natexlab\endcsname\relax\def\natexlab#1{#1}\fi
\providecommand{\url}[1]{\href{#1}{#1}}
\providecommand{\dodoi}[1]{doi:~\href{http://doi.org/#1}{\nolinkurl{#1}}}
\providecommand{\doeprint}[1]{\href{http://ascl.net/#1}{\nolinkurl{http://ascl.net/#1}}}
\providecommand{\doarXiv}[1]{\href{https://arxiv.org/abs/#1}{\nolinkurl{https://arxiv.org/abs/#1}}}

\bibitem[{{Abramowicz}(1971)}]{1971AcA....21...81Abramowicz}
{Abramowicz}, M.~A. 1971, \actaa, 21, 81

\bibitem[{{Akmal} {et~al.}(1998){Akmal}, {Pandharipande}, \& {Ravenhall}}]{1998Akmal}
{Akmal}, A., {Pandharipande}, V.~R., \& {Ravenhall}, D.~G. 1998, \prc, 58, 1804, \dodoi{10.1103/PhysRevC.58.1804}

\bibitem[{{Alford} {et~al.}(2022){Alford}, {Brodie}, {Haber}, \& {Tews}}]{QMCRMF3}
{Alford}, M.~G., {Brodie}, L., {Haber}, A., \& {Tews}, I. 2022, \prc, 106, 055804, \dodoi{10.1103/PhysRevC.106.055804}

\bibitem[{{AlGendy} \& {Morsink}(2014)}]{2014AlGendy}
{AlGendy}, M., \& {Morsink}, S.~M. 2014, ApJ, 791, 78, \dodoi{10.1088/0004-637X/791/2/78}

\bibitem[{{Arnett} \& {Bowers}(1977)}]{1977ApJS...33..415Arnett}
{Arnett}, W.~D., \& {Bowers}, R.~L. 1977, \apjs, 33, 415, \dodoi{10.1086/190434}

\bibitem[{{Baldo} {et~al.}(1997){Baldo}, {Bombaci}, \& {Burgio}}]{BBB}
{Baldo}, M., {Bombaci}, I., \& {Burgio}, G.~F. 1997, Astronomy and Astrophysics, 328, 274, \dodoi{10.48550/arXiv.astro-ph/9707277}

\bibitem[{{Bardeen} \& {Wagoner}(1971)}]{1971ApJ...167..359Bardeen}
{Bardeen}, J.~M., \& {Wagoner}, R.~V. 1971, \apj, 167, 359, \dodoi{10.1086/151039}

\bibitem[{{Baub{\"o}ck} {et~al.}(2013){Baub{\"o}ck}, {Psaltis}, \& {{\"O}zel}}]{2013ApJ...766...87Baubocka}
{Baub{\"o}ck}, M., {Psaltis}, D., \& {{\"O}zel}, F. 2013, \apj, 766, 87, \dodoi{10.1088/0004-637X/766/2/87}

\bibitem[{{Bejger}(2013)}]{2013A&A...552A..59Bejger}
{Bejger}, M. 2013, \aap, 552, A59, \dodoi{10.1051/0004-6361/201220876}

\bibitem[{{Berti} {et~al.}(2005){Berti}, {White}, {Maniopoulou}, \& {Bruni}}]{2005MNRAS.358..923Berti}
{Berti}, E., {White}, F., {Maniopoulou}, A., \& {Bruni}, M. 2005, \mnras, 358, 923, \dodoi{10.1111/j.1365-2966.2005.08812.x}

\bibitem[{{Bhattacharyya} {et~al.}(2006){Bhattacharyya}, {Miller}, \& {Lamb}}]{2006ApJ...644.1085Bhattacharyya}
{Bhattacharyya}, S., {Miller}, M.~C., \& {Lamb}, F.~K. 2006, \apj, 644, 1085, \dodoi{10.1086/503860}

\bibitem[{{Bogdanov} {et~al.}(2019){Bogdanov}, {Lamb}, {Mahmoodifar}, {Miller}, {Morsink}, {Riley}, {Strohmayer}, {Tung}, {Watts}, {Dittmann}, {Chakrabarty}, {Guillot}, {Arzoumanian}, \& {Gendreau}}]{2019bBogdanov}
{Bogdanov}, S., {Lamb}, F.~K., {Mahmoodifar}, S., {et~al.} 2019, \apjl, 887, L26, \dodoi{10.3847/2041-8213/ab5968}

\bibitem[{{Cadeau} {et~al.}(2007){Cadeau}, {Morsink}, {Leahy}, \& {Campbell}}]{2007ApJ...654..458Cadeau}
{Cadeau}, C., {Morsink}, S.~M., {Leahy}, D., \& {Campbell}, S.~S. 2007, \apj, 654, 458, \dodoi{10.1086/509103}

\bibitem[{{Chang} {et~al.}(2006){Chang}, {Morsink}, {Bildsten}, \& {Wasserman}}]{2006ApJ...636L.117Chang}
{Chang}, P., {Morsink}, S., {Bildsten}, L., \& {Wasserman}, I. 2006, \apjl, 636, L117, \dodoi{10.1086/499428}

\bibitem[{{Char} {et~al.}(2023){Char}, {Mondal}, {Gulminelli}, \& {Oertel}}]{GDFMII}
{Char}, P., {Mondal}, C., {Gulminelli}, F., \& {Oertel}, M. 2023, \prd, 108, 103045, \dodoi{10.1103/PhysRevD.108.103045}

\bibitem[{{Choudhury} {et~al.}(2024){Choudhury}, {Salmi}, {Vinciguerra}, {Riley}, {Kini}, {Watts}, {Dorsman}, {Bogdanov}, {Guillot}, {Ray}, {Reardon}, {Remillard}, {Bilous}, {Huppenkothen}, {Lattimer}, {Rutherford}, {Arzoumanian}, {Gendreau}, {Morsink}, \& {Ho}}]{2024ApJ...971L..20Choudhury}
{Choudhury}, D., {Salmi}, T., {Vinciguerra}, S., {et~al.} 2024, \apjl, 971, L20, \dodoi{10.3847/2041-8213/ad5a6f}

\bibitem[{{Cruise} {et~al.}(2025){Cruise}, {Guainazzi}, {Aird}, {Carrera}, {Costantini}, {Corrales}, {Dauser}, {Eckert}, {Gastaldello}, {Matsumoto}, {Osten}, {Petrucci}, {Porquet}, {Pratt}, {Rea}, {Reiprich}, {Simionescu}, {Spiga}, \& {Troja}}]{2025NatAs...9...36Cruise}
{Cruise}, M., {Guainazzi}, M., {Aird}, J., {et~al.} 2025, Nature Astronomy, 9, 36, \dodoi{10.1038/s41550-024-02416-3}

\bibitem[{{Dexheimer} \& {Schramm}(2008)}]{CMF1}
{Dexheimer}, V., \& {Schramm}, S. 2008, The Astrophysical Journal, 683, 943, \dodoi{10.1086/589735}

\bibitem[{{Dittmann} {et~al.}(2024){Dittmann}, {Miller}, {Lamb}, {Holt}, {Chirenti}, {Wolff}, {Bogdanov}, {Guillot}, {Ho}, {Morsink}, {Arzoumanian}, \& {Gendreau}}]{2024ApJ...974..295Dittman}
{Dittmann}, A.~J., {Miller}, M.~C., {Lamb}, F.~K., {et~al.} 2024, \apj, 974, 295, \dodoi{10.3847/1538-4357/ad5f1e}

\bibitem[{{Dorsman} {et~al.}(2025){Dorsman}, {Salmi}, {Watts}, {Ng}, {Kamath}, {Bobrikova}, {Poutanen}, {Loktev}, {Kini}, {Choudhury}, {Vinciguerra}, {Bogdanov}, \& {Chakrabarty}}]{2025MNRAS.538.2853Dorsman}
{Dorsman}, B., {Salmi}, T., {Watts}, A.~L., {et~al.} 2025, \mnras, 538, 2853, \dodoi{10.1093/mnras/staf438}

\bibitem[{{Fonseca} {et~al.}(2021){Fonseca}, {Cromartie}, {Pennucci}, {Ray}, {Kirichenko}, {Ransom}, {Demorest}, {Stairs}, {Arzoumanian}, {Guillemot}, {Parthasarathy}, {Kerr}, {Cognard}, {Baker}, {Blumer}, {Brook}, {DeCesar}, {Dolch}, {Dong}, {Ferrara}, {Fiore}, {Garver-Daniels}, {Good}, {Jennings}, {Jones}, {Kaspi}, {Lam}, {Lorimer}, {Luo}, {McEwen}, {McKee}, {McLaughlin}, {McMann}, {Meyers}, {Naidu}, {Ng}, {Nice}, {Pol}, {Radovan}, {Shapiro-Albert}, {Tan}, {Tendulkar}, {Swiggum}, {Wahl}, \& {Zhu}}]{2021ApJ...915L..12Fonseca}
{Fonseca}, E., {Cromartie}, H.~T., {Pennucci}, T.~T., {et~al.} 2021, \apjl, 915, L12, \dodoi{10.3847/2041-8213/ac03b8}

\bibitem[{{Friedman} {et~al.}(1986){Friedman}, {Ipser}, \& {Parker}}]{1986ApJ...304..115Friedman}
{Friedman}, J.~L., {Ipser}, J.~R., \& {Parker}, L. 1986, \apj, 304, 115, \dodoi{10.1086/164149}

\bibitem[{{Friedman} \& {Stergioulas}(2013)}]{2013rrs..book.....Friedman}
{Friedman}, J.~L., \& {Stergioulas}, N. 2013, {Rotating Relativistic Stars} (Cambridge University Press)

\bibitem[{{Glampedakis} \& {Babak}(2006)}]{2006CQGra..23.4167Glampedakis}
{Glampedakis}, K., \& {Babak}, S. 2006, Classical and Quantum Gravity, 23, 4167, \dodoi{10.1088/0264-9381/23/12/013}

\bibitem[{{Gulminelli} \& {Raduta}(2015)}]{KDE0v-Rs-SK255-SKa}
{Gulminelli}, F., \& {Raduta}, A.~R. 2015, \prc, 92, 055803, \dodoi{10.1103/PhysRevC.92.055803}

\bibitem[{{Hartle}(1967)}]{1967Hartle}
{Hartle}, J.~B. 1967, \apj, 150, 1005, \dodoi{10.1086/149400}

\bibitem[{{Hartle} \& {Thorne}(1968)}]{1968Hartle}
{Hartle}, J.~B., \& {Thorne}, K.~S. 1968, \apj, 153, 807, \dodoi{10.1086/149707}

\bibitem[{{Hebeler} {et~al.}(2013){Hebeler}, {Lattimer}, {Pethick}, \& {Schwenk}}]{hlps-paper}
{Hebeler}, K., {Lattimer}, J.~M., {Pethick}, C.~J., \& {Schwenk}, A. 2013, ApJ, 773, 11, \dodoi{10.1088/0004-637X/773/1/11}

\bibitem[{{Kini} {et~al.}(2024){Kini}, {Salmi}, {Vinciguerra}, {Watts}, {Bilous}, {Galloway}, {van der Wateren}, {Khalsa}, {Bogdanov}, {Buchner}, \& {Suleimanov}}]{2024MNRAS.535.1507Kini}
{Kini}, Y., {Salmi}, T., {Vinciguerra}, S., {et~al.} 2024, \mnras, 535, 1507, \dodoi{10.1093/mnras/stae2398}

\bibitem[{{Kini} {et~al.}(2025){Kini}, {Watts}, {Salmi}, {Bilous}, {Vinciguerra}, {Guillot}, {Ballantyne}, {Kuulkers}, {Bogdanov}, \& {Suleimanov}}]{2025MNRAS.tmp..892Kini}
{Kini}, Y., {Watts}, A.~L., {Salmi}, T., {et~al.} 2025, \mnras, \dodoi{10.1093/mnras/staf908}

\bibitem[{{Konstantinou} \& {Morsink}(2022)}]{2022ApJ...934..139Konstantinou}
{Konstantinou}, A., \& {Morsink}, S.~M. 2022, \apj, 934, 139, \dodoi{10.3847/1538-4357/ac7b86}

\bibitem[{{Laarakkers} \& {Poisson}(1999)}]{1999ApJ...512..282Laarakkers}
{Laarakkers}, W.~G., \& {Poisson}, E. 1999, \apj, 512, 282, \dodoi{10.1086/306732}

\bibitem[{{Li} {et~al.}(2025){Li}, {Watts}, {Zhang}, {Guillot}, {Xu}, {Santangelo}, {Zane}, {Feng}, {Zhang}, {Ge}, {Qi}, {Salmi}, {Dorsman}, {Miao}, {Tu}, {Cavecchi}, {Zhou}, {Zheng}, {Wang}, {Cheng}, {Liu}, {Wei}, {Wang}, {Xu}, {Weng}, {Zhu}, {Li}, {Shao}, {Tuo}, {Dohi}, {Lyu}, {Liu}, {Yuan}, {Wang}, {Zhang}, {Li}, {Tao}, {Zhang}, {Shen}, {Provid{\^e}ncia}, {Tolos}, {Patruno}, {Li}, {Liu}, {Zhou}, {Chen}, {Fan}, {Kajino}, {Lai}, {Li}, {Meng}, {Tang}, {Xiao}, {Xiong}, {Xu}, {Zhou}, {Ballantyne}, {Fiorella Burgio}, {Chenevez}, {Choudhury}, {Fantina}, {Galloway}, {Gulminelli}, {Hebeler}, {Hoogkamer}, {Kini}, {Kurkela}, {Linares}, {Margueron}, {Mendes}, {Oertel}, {Papitto}, {Poutanen}, {Rea}, {Schwenk}, {Svensson}, {Tsang}, {Vuorinen}, {Andersson}, {Miller}, {Rezzolla}, {Stone}, \& {Thomas}}]{2025arXiv250608104Li}
{Li}, A., {Watts}, A.~L., {Zhang}, G., {et~al.} 2025, arXiv e-prints, arXiv:2506.08104, \dodoi{10.48550/arXiv.2506.08104}

\bibitem[{{Lin} {et~al.}(2010){Lin}, {{\"O}zel}, {Chakrabarty}, \& {Psaltis}}]{2010ApJ...723.1053Lin}
{Lin}, J., {{\"O}zel}, F., {Chakrabarty}, D., \& {Psaltis}, D. 2010, \apj, 723, 1053, \dodoi{10.1088/0004-637X/723/2/1053}

\bibitem[{{Lindblom}(1992)}]{1992Lindblom}
{Lindblom}, L. 1992, \apj, 398, 569, \dodoi{10.1086/171882}

\bibitem[{{Logoteta} {et~al.}(2018){Logoteta}, {Bombaci}, \& {Kievsky}}]{BLC}
{Logoteta}, D., {Bombaci}, I., \& {Kievsky}, A. 2018, in Journal of Physics Conference Series, Vol. 981, Journal of Physics Conference Series (IOP), 012009, \dodoi{10.1088/1742-6596/981/1/012009}

\bibitem[{{Majumder} {et~al.}(2015){Majumder}, {Yagi}, \& {Yunes}}]{2015PhRvD..92b4020Majumder}
{Majumder}, B., {Yagi}, K., \& {Yunes}, N. 2015, \prd, 92, 024020, \dodoi{10.1103/PhysRevD.92.024020}

\bibitem[{{Mauviard} {et~al.}(2025){Mauviard}, {Guillot}, {Salmi}, {Choudhury}, {Dorsman}, {Gonz{\'a}lez-Caniulef}, {Hoogkamer}, {Huppenkothen}, {Kazantsev}, {Kini}, {Olive}, {Stammler}, {Watts}, {Mendes}, {Rutherford}, {Schwenk}, {Svensson}, {Bogdanov}, {Kerr}, {Ray}, {Guillemot}, {Cognard}, \& {Theureau}}]{2025arXiv250614883Mauviard}
{Mauviard}, L., {Guillot}, S., {Salmi}, T., {et~al.} 2025, arXiv e-prints, arXiv:2506.14883, \dodoi{10.48550/arXiv.2506.14883}

\bibitem[{{Miller} \& {Lamb}(1998)}]{1998ApJ...499L..37Miller}
{Miller}, M.~C., \& {Lamb}, F.~K. 1998, \apjl, 499, L37, \dodoi{10.1086/311335}

\bibitem[{{Miller} {et~al.}(2019){Miller}, {Lamb}, {Dittmann}, {Bogdanov}, {Arzoumanian}, {Gendreau}, {Guillot}, {Harding}, {Ho}, {Lattimer}, {Ludlam}, {Mahmoodifar}, {Morsink}, {Ray}, {Strohmayer}, {Wood}, {Enoto}, {Foster}, {Okajima}, {Prigozhin}, \& {Soong}}]{2019Miller}
{Miller}, M.~C., {Lamb}, F.~K., {Dittmann}, A.~J., {et~al.} 2019, \apjl, 887, L24, \dodoi{10.3847/2041-8213/ab50c5}

\bibitem[{{Mondal} {et~al.}(2020){Mondal}, {Vi{\~n}as}, {Centelles}, \& {De}}]{D1MStar}
{Mondal}, C., {Vi{\~n}as}, X., {Centelles}, M., \& {De}, J.~N. 2020, \prc, 102, 015802, \dodoi{10.1103/PhysRevC.102.015802}

\bibitem[{{Morsink} {et~al.}(2007){Morsink}, {Leahy}, {Cadeau}, \& {Braga}}]{2007Morsink}
{Morsink}, S.~M., {Leahy}, D.~A., {Cadeau}, C., \& {Braga}, J. 2007, ApJ, 663, 1244, \dodoi{10.1086/518648}

\bibitem[{{N{\"a}ttil{\"a}} {et~al.}(2017){N{\"a}ttil{\"a}}, {Miller}, {Steiner}, {Kajava}, {Suleimanov}, \& {Poutanen}}]{2017A&A...608A..31Nattila}
{N{\"a}ttil{\"a}}, J., {Miller}, M.~C., {Steiner}, A.~W., {et~al.} 2017, \aap, 608, A31, \dodoi{10.1051/0004-6361/201731082}

\bibitem[{{N{\"a}ttil{\"a}} \& {Pihajoki}(2018)}]{2018A&A...615A..50Nattila}
{N{\"a}ttil{\"a}}, J., \& {Pihajoki}, P. 2018, \aap, 615, A50, \dodoi{10.1051/0004-6361/201630261}

\bibitem[{{Oliva} \& {Frutos-Alfaro}(2021)}]{2021MNRAS.505.2870Oliva}
{Oliva}, G.~A., \& {Frutos-Alfaro}, F. 2021, \mnras, 505, 2870, \dodoi{10.1093/mnras/stab1380}

\bibitem[{{{\"O}zel} \& {Psaltis}(2003)}]{2003ApJ...582L..31Ozel}
{{\"O}zel}, F., \& {Psaltis}, D. 2003, \apjl, 582, L31, \dodoi{10.1086/346197}

\bibitem[{{Papigkiotis} {et~al.}(2025){Papigkiotis}, {Vardakas}, {Likas}, \& {Stergioulas}}]{2025PhRvD.111h3056Papigkiotis}
{Papigkiotis}, G., {Vardakas}, G., {Likas}, A., \& {Stergioulas}, N. 2025, \prd, 111, 083056, \dodoi{10.1103/PhysRevD.111.083056}

\bibitem[{{Pappas} \& {Apostolatos}(2012)}]{2012PhRvL.108w1104Pappas}
{Pappas}, G., \& {Apostolatos}, T.~A. 2012, \prl, 108, 231104, \dodoi{10.1103/PhysRevLett.108.231104}

\bibitem[{{Pappas} \& {Apostolatos}(2014)}]{2014PhRvL.112l1101Pappas}
---. 2014, \prl, 112, 121101, \dodoi{10.1103/PhysRevLett.112.121101}

\bibitem[{{Pearson} {et~al.}(2018){Pearson}, {Chamel}, {Potekhin}, {Fantina}, {Ducoin}, {Dutta}, \& {Goriely}}]{M3-TW}
{Pearson}, J.~M., {Chamel}, N., {Potekhin}, A.~Y., {et~al.} 2018, Monthly Notices of the Royal Astronomical Society, 481, 2994, \dodoi{10.1093/mnras/sty2413}

\bibitem[{{Pihajoki} {et~al.}(2018){Pihajoki}, {Mannerkoski}, {N{\"a}ttil{\"a}}, \& {Johansson}}]{2018ApJ...863....8Pihajoki}
{Pihajoki}, P., {Mannerkoski}, M., {N{\"a}ttil{\"a}}, J., \& {Johansson}, P.~H. 2018, \apj, 863, 8, \dodoi{10.3847/1538-4357/aacea0}

\bibitem[{{Poutanen} \& {Gierli{\'n}ski}(2003)}]{2003Poutanen}
{Poutanen}, J., \& {Gierli{\'n}ski}, M. 2003, \mnras, 343, 1301, \dodoi{10.1046/j.1365-8711.2003.06773.x}

\bibitem[{{Pradhan} {et~al.}(2023){Pradhan}, {Chatterjee}, {Gandhi}, \& {Schaffner-Bielich}}]{PCSB2}
{Pradhan}, B.~K., {Chatterjee}, D., {Gandhi}, R., \& {Schaffner-Bielich}, J. 2023, Nuclear Physics, Section A, 1030, 122578, \dodoi{10.1016/j.nuclphysa.2022.122578}

\bibitem[{Read {et~al.}(2009)Read, Lackey, Owen, \& Friedman}]{Read2009}
Read, J.~S., Lackey, B.~D., Owen, B.~J., \& Friedman, J.~L. 2009, \prd, 79, \dodoi{10.1103/physrevd.79.124032}

\bibitem[{{Riley} {et~al.}(2019){Riley}, {Watts}, {Bogdanov}, {Ray}, {Ludlam}, {Guillot}, {Arzoumanian}, {Baker}, {Bilous}, {Chakrabarty}, {Gendreau}, {Harding}, {Ho}, {Lattimer}, {Morsink}, \& {Strohmayer}}]{2019Riley}
{Riley}, T.~E., {Watts}, A.~L., {Bogdanov}, S., {et~al.} 2019, \apjl, 887, L21, \dodoi{10.3847/2041-8213/ab481c}

\bibitem[{{Salmi} {et~al.}(2024{\natexlab{a}}){Salmi}, {Deneva}, {Ray}, {Watts}, {Choudhury}, {Kini}, {Vinciguerra}, {Cromartie}, {Wolff}, {Arzoumanian}, {Bogdanov}, {Gendreau}, {Guillot}, {Ho}, {Morsink}, {Cognard}, {Guillemot}, {Theureau}, \& {Kerr}}]{2024ApJ...976...58Salmi}
{Salmi}, T., {Deneva}, J.~S., {Ray}, P.~S., {et~al.} 2024{\natexlab{a}}, \apj, 976, 58, \dodoi{10.3847/1538-4357/ad81d2}

\bibitem[{{Salmi} {et~al.}(2024{\natexlab{b}}){Salmi}, {Choudhury}, {Kini}, {Riley}, {Vinciguerra}, {Watts}, {Wolff}, {Arzoumanian}, {Bogdanov}, {Chakrabarty}, {Gendreau}, {Guillot}, {Ho}, {Huppenkothen}, {Ludlam}, {Morsink}, \& {Ray}}]{2024ApJ...974..294Salmi}
{Salmi}, T., {Choudhury}, D., {Kini}, Y., {et~al.} 2024{\natexlab{b}}, \apj, 974, 294, \dodoi{10.3847/1538-4357/ad5f1f}

\bibitem[{{Silva} {et~al.}(2021){Silva}, {Pappas}, {Yunes}, \& {Yagi}}]{2021PhRvD.103f3038Silva}
{Silva}, H.~O., {Pappas}, G., {Yunes}, N., \& {Yagi}, K. 2021, \prd, 103, 063038, \dodoi{10.1103/PhysRevD.103.063038}

\bibitem[{{Steiner} {et~al.}(2018){Steiner}, {Heinke}, {Bogdanov}, {Li}, {Ho}, {Bahramian}, \& {Han}}]{2018Steiner}
{Steiner}, A.~W., {Heinke}, C.~O., {Bogdanov}, S., {et~al.} 2018, \mnras, 476, 421, \dodoi{10.1093/mnras/sty215}

\bibitem[{{Stergioulas} \& {Friedman}(1995)}]{1995Stergioulas}
{Stergioulas}, N., \& {Friedman}, J.~L. 1995, ApJ, 444, 306, \dodoi{10.1086/175605}

\bibitem[{Storn \& Price(1997)}]{DiffEv}
Storn, R., \& Price, K.~V. 1997, Journal of Global Optimization, 11, 341.
\newblock \url{https://api.semanticscholar.org/CorpusID:5297867}

\bibitem[{{Togashi} {et~al.}(2017){Togashi}, {Nakazato}, {Takehara}, {Yamamuro}, {Suzuki}, \& {Takano}}]{QHC21A}
{Togashi}, H., {Nakazato}, K., {Takehara}, Y., {et~al.} 2017, Nuclear Physics, Section A, 961, 78, \dodoi{10.1016/j.nuclphysa.2017.02.010}

\bibitem[{{Typel} {et~al.}(2015){Typel}, {Oertel}, \& {Kl{\"a}hn}}]{2015PPN....46..633Typel}
{Typel}, S., {Oertel}, M., \& {Kl{\"a}hn}, T. 2015, Physics of Particles and Nuclei, 46, 633, \dodoi{10.1134/S1063779615040061}

\bibitem[{{Typel} \& {Wolter}(1999)}]{1999NuPhA.656..331Typel}
{Typel}, S., \& {Wolter}, H.~H. 1999, \nphysa, 656, 331, \dodoi{10.1016/S0375-9474(99)00310-3}

\bibitem[{Virtanen {et~al.}(2020)Virtanen, Gommers, Oliphant, Haberland, Reddy, Cournapeau, Burovski, Peterson, Weckesser, Bright, {van der Walt}, Brett, Wilson, Millman, Mayorov, Nelson, Jones, Kern, Larson, Carey, Polat, Feng, Moore, {VanderPlas}, Laxalde, Perktold, Cimrman, Henriksen, Quintero, Harris, Archibald, Ribeiro, Pedregosa, {van Mulbregt}, \& {SciPy 1.0 Contributors}}]{2020SciPy-NMeth}
Virtanen, P., Gommers, R., Oliphant, T.~E., {et~al.} 2020, Nature Methods, 17, 261, \dodoi{10.1038/s41592-019-0686-2}

\bibitem[{{Watts}(2012)}]{2012Watts}
{Watts}, A.~L. 2012, \araa, 50, 609, \dodoi{10.1146/annurev-astro-040312-132617}

\bibitem[{{Xia} {et~al.}(2022){Xia}, {Maruyama}, {Li}, {Yuan Sun}, {Long}, \& {Zhang}}]{DDMEX-NL3-PKDD}
{Xia}, C.-J., {Maruyama}, T., {Li}, A., {et~al.} 2022, Communications in Theoretical Physics, 74, 095303, \dodoi{10.1088/1572-9494/ac71fd}

\bibitem[{{Yagi} {et~al.}(2014){Yagi}, {Kyutoku}, {Pappas}, {Yunes}, \& {Apostolatos}}]{2014PhRvD..89l4013Yagi}
{Yagi}, K., {Kyutoku}, K., {Pappas}, G., {Yunes}, N., \& {Apostolatos}, T.~A. 2014, \prd, 89, 124013, \dodoi{10.1103/PhysRevD.89.124013}

\end{thebibliography}
\bibliographystyle{aasjournal}



\end{document}